\documentstyle[preprint,aps,epsfig]{revtex}
\title{Spin-Orbit-Induced Magnetic Anisotropy for Impurities in 
Metallic Samples II. Finite Size Dependence in the Kondo Resistivity}
\author{O. \'Ujs\'aghy$^a$ and A. Zawadowski$^{a,b}$} 
\address{$^a$Institute of Physics and Research Group of Hungarian Academy of
Sciences, Technical University of Budapest, H-1521 Budapest, Hungary}
\address{$^b$Research Institute for Solid State Physics, POB 49, H-1525
Budapest, Hungary}
\date{\today}
\begin{document}

\draft
\maketitle

\begin{abstract}
The electrical resistivity including the Kondo resistivity increase at low 
temperature is calculated for thin films of dilute magnetic
alloys. Assuming that in the non-magnetic host the spin-orbit interaction is
strong like in Au and Cu, the magnetic impurities have a surface anisotropy
calculated in part I. That anisotropy hinders the motion of the
spin. Including that anisotropy the effective electron-impurity
coupling is calculated by using the second order renormalization
group equations. The amplitude of the Kondo resistivity
contribution is reduced as the position of the impurity
approaches the surface but the increase occurs approximately at
the bulk Kondo temperature.
Different proximity effects observed by Giordano 
are also explained qualitatively
where the films of magnetic alloys are covered by pure second
films with different mean free path. The theory explains the
experimental results in those cases where a considerable amount
of impurities is at the surface inside the ballistic region.
\end{abstract}

\section{Introduction}
\label{sec:1}

In the previous paper \cite{Ujsaghy1} (Part I) following Ref's 
\cite{Ujsaghy} and \cite{proc} we have calculated the 
magnetic anisotropy for a magnetic impurity embodied into a non-magnetic
host (e.g. Au, Cu) with large spin-orbit interaction for the conduction 
electrons on the sites of the host atoms. The magnetic anisotropy is developed 
due to the exchange interaction between the magnetic impurities and the 
conduction electrons. As the scattering due to impurities involves different 
angular momentum channels ($-l<m<l$), therefore, the scattering depends on 
the directions of the conduction electrons before and after the scattering.
In that way the scattering on the impurity itself depends on which host atoms
the electrons are scattered by the spin-orbit interaction. Due to that dependence 
the anisotropy is determined by the positions of host atoms around the
impurity and it is the larger the stronger the asymmetry around the
impurity. As that information is carried by the momenta, therefore, 
that is restricted to the atoms in the range of the elastic
mean free path $l_{\text{el}}$. Thus the anisotropy can be developed
only if the impurity is inside surface area of the thickness of the
elastic mean free path (ballistic region of the surface). If the surface 
in that region is plane like then the anisotropy energy is
(see Eq.\ (1) of Part I \cite{Ujsaghy1})
\begin{equation}
H_a=K_d ({\bf n}{\bf S})^2,
\label{aniz}\end{equation}
where ${\bf S}$ is the spin operator of the impurity, ${\bf n}$ is the
normal direction of the experienced surface element, and $K_d$ is the 
anisotropy constant depending apart from the oscillatory part like
$1/d$ on the distance $d$, measured from the surface. 
Of course, if the ballistic
region contains more sophisticated part of the surface, then the determination
of the direction and amplitude of the anisotropy is a more complex task.
The oscillatory part decays faster than $1/d$ approaching the bulk part 
of the sample (see Eq.\ (B20b) in Part I).

\begin{figure}
  \begin{center}
    \epsfig{file=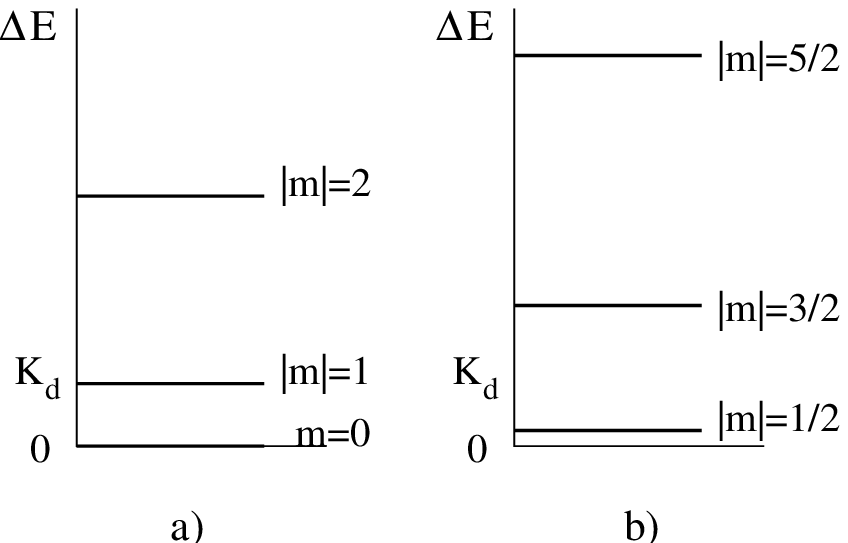}
  \end{center}
\caption{The splitting schema due to the anisotropy for (a) $S=2$ and
(b) $S=5/2$.}
\label{fig01}
\end{figure}

\begin{figure}
  \begin{center}
    \epsfig{file=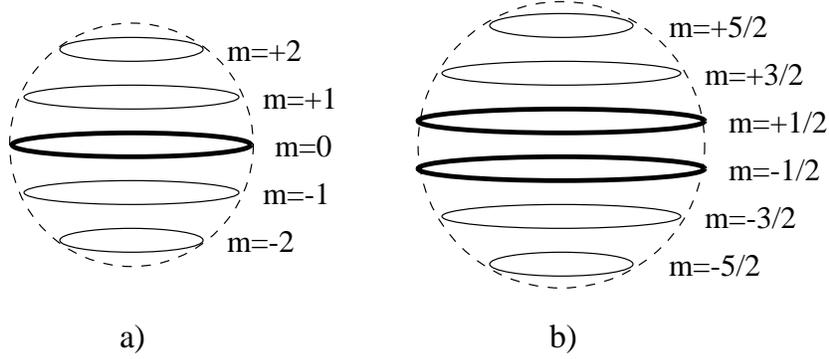}
  \end{center}
\caption{The squizing of the spin into planar states due to the anisotropy is
illustrated for (a) $S=2$ and (b) $S=5/2$. The lowest energy states
are shown by heavy lines.}
\label{fig02}
\end{figure}

In the surface area the anisotropy energy leads to different splitting schema
shown in Fig.\ \ref{fig01}, depending whether the spin $S$ is integer or 
half-integer. 
In this way the spin very nearby the surface is freezing into a singlet or
doublet considering the integer and half-integer case, respectively.
Thus at low enough temperature the spin shows no, or restricted dynamics.
It is important to point out, that the states $m=\pm 1/2$ for $S=5/2$ cannot
be replaced by a doublet of $S=1/2$. The spins are rather squeezed into planar 
states as shown in Fig.\ \ref{fig02}.

\begin{figure}
  \begin{center}
    \epsfig{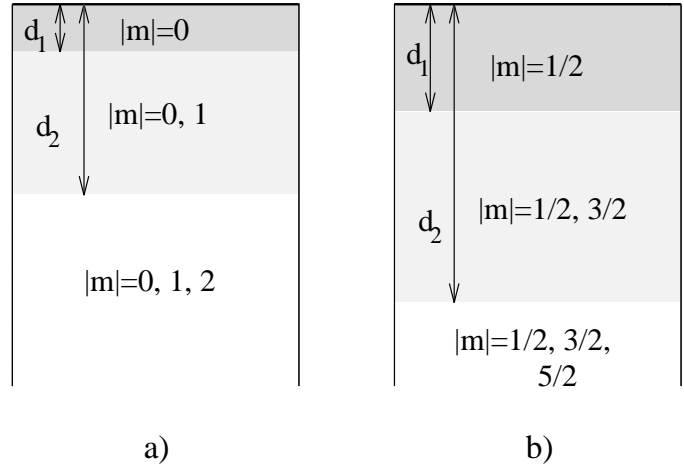}
  \end{center}
\caption{The different layers of the impurity positions where
more orbitals become
active as the impurity positions approach the bulk assuming that
$d_1, d_2<l_{\text{el}}$. (a) $S=2$, (b) $S=5/2$.}
\label{fig03}
\end{figure}

Assuming that $K_d\gg T$, the integer spin does not contributes to the 
resistivity in contrary with the case of the half-integer spin where 
the two lowest states contributes to the resistivity.
The spin $S=1/2$ is, however, not affected by anisotropy.
Thus different behaviors of the electrical resistivity
can be expected depending on the value of the spin. Considering
impurities nearby the surface
inside the ballistic region more and more 
orbitals become active as the impurity positions approaching the bulk
(see Fig.\ \ref{fig03}). 

Similar structure appears in the temperature dependence
of the resistivity. Cooling down the sample, at the beginning almost all the
spins are free. At further cooling more and more spin states are frozen
in thus in case $S=2$ at $T<T_1$ three orbitals and $T<T_2$ a single orbital 
is populated ($T_1:T_2=4:1$) while in case of $S=5/2$ each state
is double degenerate ($T_1:T_2=25:9$).

\begin{figure}
  \begin{center}
    \epsfig{file=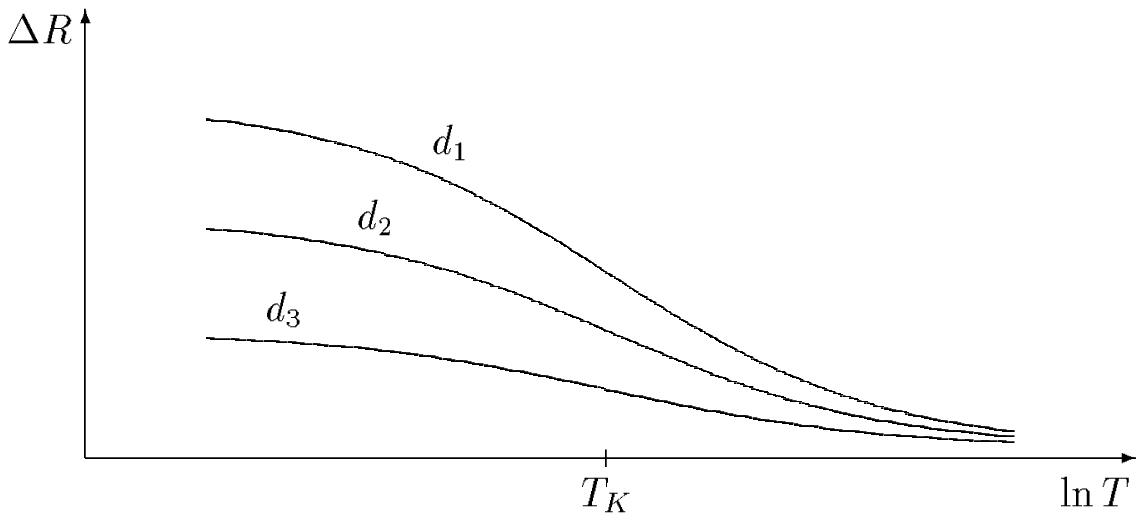}
  \end{center}
\caption{The schematic plots of the contribution to the Kondo resistivity
for impurities with different distances $d_3<d_2<d_1$
measured from the surface.}
\label{fig04}
\end{figure}

Assuming that in the region of the Kondo temperature $T_K$ the occupations of
the different states are varying by considerable amount 
depending on the positions of the impurities in the surface region,
then lowering
the temperature less and less impurities can further develop the Kondo
state, thus less and less impurities can contribute to further
increase of the Kondo resistivity. As it has
already been pointed out the reduction in the contribution to resistivity
somewhat less pronounced for half-integer spin than for integer spin 
(see the discussion of Fig.\ \ref{fig02}). The contribution to
the Kondo resistivity can be 
schematically plotted for impurities with different distances measured from
the surface $d_3<d_2<d_1$ (see Fig.~\ref{fig04}).

The phenomena described above are very similar to the Kondo effect in the
presence of crystalline field at the impurity site when lowering the
temperature different crystal field states are frozen out, but those
fields are identical for all impurities of the same kind. Such calculations
have been performed e.g. by Shin-ichi Kashiba {\it et al.} \cite{Kashiba}.

The reduction in the averaged Kondo resistivity is sensitive on
the size  of the 
sample e.g. film thickness or diameter of the wire etc. as the ratio of the 
surface influenced impurities to the total number of the impurities goes to 
zero as the size is further increased. The role of the surface
on the impurities can be reduced by depositing another pure film
on the surface of the 
sample. The effect can be also influenced by changing the elastic mean free 
path in the samples or in the deposited films. The effects of
those kinds will be 
summarized and discussed at the end of the paper (see Sec.~\ref{sec:7})
making use the qualitative results obtained for the resistivity in
Sec.~\ref{sec:5} and \ref{sec:6} and all of the references can be found there.

For the actual calculation of the resistivity first the temperature 
dependence of the effective exchange coupling must be calculated.
For that, except the very low temperature region, the second order
multiplicative renormalization group transformations (two-loops
approximation) can be used which give smooth behavior at the Kondo 
temperature in contrast to the first order scaling (one-loop
approximation) which results in an artificial divergence at the Kondo 
temperature. Those methods are generalized by taking into account the
surface anisotropy terms which occur as a low energy cutoff of the 
logarithmic
integrals in the calculation of certain diagrams. The different diagrams 
depending on their spin labels have different infrared, low energy cutoff
due to the anisotropy. These calculations are in close analogy to those
with crystalline splitting.
The next step is using these effective couplings to calculate the electrical
resistivity by solving the Boltzmann equation and an average over
the impurity positions is taken also.

Finally, to fit the calculated resistivity at low temperature, an effective
surface layer thickness $\lambda$ can be introduced, by assuming that
inside of that surface region there is no Kondo effect, and outside the
Kondo anomaly is fully developed. The experimental data are compared both 
with that phenomenological description and the original calculations and 
they are giving equally excellent fit.

The paper is organized as follows. In Sec.~\ref{sec:2} the general scheme
of the multiplicative renormalization group (MRG) is presented for the
Hamiltonian with the anisotropy term. The scaling equations are presented in 
Sec.~\ref{sec:3} which are solved in Sec.~\ref{sec:4}. The electrical
resistivity contribution is calculated in Sec.~\ref{sec:5} for the dilute
limit and Kondo resistivity in thin films is given in Sec.~\ref{sec:6}.
Sec.~\ref{sec:7} is devoted to the experimental results, theoretical
interpretation of the results and prediction. A general discussion is contained
in Sec.~\ref{sec:8}. The Appendix contains the actual calculation
of the diagrams which are needed in Sec.~\ref{sec:3}.
Throughout the paper the $\hbar=k_B=1$ units were used.
 
\section{The Hamiltonian and the general form of the MRG transformation}
\label{sec:2}

The Kondo Hamiltonian in the presence of the anisotropy is
\begin{eqnarray}
H&=&\sum\limits_{k,\sigma}\!\!\varepsilon_k \,a_{k\sigma}^\dagger
a_{k\sigma}+H_a\nonumber \\
&+& \sum\limits_{\scriptstyle k,k^\prime,\sigma,\sigma^\prime \atop
\scriptstyle M,M^\prime}
J_{MM'} \bbox{S}_{MM'}\,(a_{k\sigma}^\dagger \bbox{\sigma}_{\sigma\sigma'}
a_{k'\sigma'}),
\label{H}\end{eqnarray}
where $a_{k\sigma}^\dagger$ ($a_{k\sigma}$) creates (annihilates)
a conduction electron with momentum $k$, spin $\sigma$ and energy 
$\varepsilon_k$ measured from the Fermi level. The conduction electron band
is taken with constant energy density $\rho_0$ for one spin direction,
with a sharp and symmetric bandwidth cutoff $D$.
$\bbox{\sigma}$ stands for the Pauli matrices,
$J_{MM'}$'s are the effective Kondo couplings and $H_a$ is given by 
Eq.\ (\ref{aniz}). 
For the impurity spin the Abrikosov' pseudofermion representation 
\cite{Abrikosov} was used
\begin{equation}
\bbox{S}=b^{\dagger}_M \bbox{S}_{MM'}b_{M'},
\end{equation}
where the projections of the $z$ component of the impurity spin 
are described by an auxiliary fermionic field $b_M$ ($M=-S\dots S$). 
Choosing the quantization axis parallel to ${\bf n}$, with this substitution 
the Hamiltonian Eq.\ (\ref{H}) become
\begin{eqnarray}
H&=&\sum\limits_{k,\sigma}\!\!\varepsilon_k \,a_{k\sigma}^\dagger
a_{k\sigma}+\sum\limits_M (\lambda+K_M) b^{\dagger}_M b_M\nonumber \\
&+&\sum\limits_{\scriptstyle k,k^\prime,\sigma,\sigma^\prime \atop
\scriptstyle M,M^\prime}
J_{MM'}(b_M^\dagger \bbox{S}_{MM'} b_{M'})(a_{k\sigma}^\dagger 
\bbox{\sigma}_{\sigma\sigma'}a_{k'\sigma'}),
\label{Hapr}\end{eqnarray}
where the chemical potential $\lambda\rightarrow\infty$ was introduced 
to project out the physical pseudofermion subspace 
$\sum\limits_M b^\dagger_M b_M=1$ and the notation $K_M=K M^2$ was introduced
for the MRG calculation.

The conduction electron and pseudofermion Green's functions are
\begin{equation}
G_{k\sigma,k'\sigma'}(\omega)={\delta_{kk'}\delta_{\sigma\sigma'}
\over \omega-\varepsilon_k-\Sigma_e}
\label{G}\end{equation}
and
\begin{equation}
{\cal G}_{MM'}({\tilde\omega})={\delta_{MM'}\over {\tilde\omega}-\lambda
-\Sigma_M({\tilde\omega})}
\label{pG}\end{equation}
where ${\tilde\omega}=\omega-K_M$.
$\Sigma_e$ and $\Sigma_M({\tilde\omega})$ are the self-energies for the 
conduction electrons and the pseudofermions, respectively. They are 
diagonal in the adequate spin quantum numbers, because of that the whole 
Hamiltonian is symmetric under rotation around the $z$ axis.
The vertex function is denoted by
$\Gamma_{k\sigma M,k'\sigma' M'}(\omega_1, \omega_2, \omega_3, \omega_4)$.

The multiplicative renormalization group transformation can be written
as \cite{Zawa}
\begin{mathletters}
\begin{equation}
G_{k\sigma,k'\sigma'}(\omega,J_{MM'}(D),K_M(D),D)=Z\biggl ({D_0\over D}
\biggr )
G_{k\sigma,k'\sigma'}(\omega,J_{MM'}^0,K_M^0,D_0)
\end{equation}
\begin{equation}
{\cal G}_{MM'}({\tilde\omega},J_{{\tilde M}{\tilde M}'}(D),K_{\tilde M}(D),D)=
Z_M\biggl ({D_0\over D}\biggr ){\cal G}_{MM'}({\tilde\omega},J_{{\tilde M}
{\tilde M}'}^0, K_{\tilde M}^0,D_0)
\end{equation}
\begin{equation}
\Gamma_{k\sigma M,k'\sigma' M'}(\omega_i,J_{{\tilde M}{\tilde M}'}(D),
K_{\tilde M}(D),D)={\Gamma_{k\sigma M,k'\sigma' M'}
(\omega_i,J_{{\tilde M}{\tilde M}'}^0,
K_{\tilde M}^0,D_0)\over Z({D_0\over D})\sqrt{ 
Z_M({D_0\over D})}\sqrt{Z_{M'}({D_0\over D})}},
\end{equation}
\end{mathletters}
where $Z\bigl({D_0\over D}\bigr)$ and $Z_M\bigl({D_0\over D}\bigr)$ are 
the renormalization factors for the electrons and pseudofermions, 
respectively. Introducing $x=\ln({D_0\over D})$ as scaling parameter, the 
Callan-Symantzik MRG equations are 
\begin{mathletters}
\begin{equation}
-\eta G+{\partial G\over\partial x}+\sum\limits_{MM'}\beta_{MM'}
{\partial G\over\partial J_{MM'}}+\sum\limits_M\gamma_M
{\partial G\over\partial K_M}=0
\end{equation}
\begin{equation}
-\eta_M {\cal G}_M+{\partial {\cal G}_M\over\partial x}+
\sum\limits_{{\tilde M}{\tilde M}'}\beta_{{\tilde M}{\tilde M}'}
{\partial {\cal G}_M\over\partial J_{{\tilde M}{\tilde M}'}}+
\sum\limits_{\tilde M}\gamma_{\tilde M}
{\partial {\cal G}_M\over\partial K_{\tilde M}}=0
\end{equation}
\begin{equation}
(\eta+{1\over 2}\eta_M+{1\over 2}\eta_{M'})\Gamma_{MM'}+
{\partial \Gamma_{MM'}\over\partial x}+
\sum\limits_{{\tilde M}{\tilde M}'}\beta_{{\tilde M}{\tilde M}'}
{\partial \Gamma_{MM'}\over\partial J_{{\tilde M}{\tilde M}'}}+
\sum\limits_{\tilde M}\gamma_{\tilde M}
{\partial \Gamma_{MM'}\over\partial K_{\tilde M}}=0,
\label{MRGc}
\end{equation}
\end{mathletters}
where 
\begin{mathletters}
\begin{equation}
\eta={d\ln Z\over dx}
\end{equation}
\begin{equation}
\eta_M={d\ln Z_M\over dx}
\end{equation}
\begin{equation}
\beta_{MM'}={dJ_{MM'}\over dx}
\end{equation}
\begin{equation}
\gamma_M={dK_M\over dx}
\end{equation}
\end{mathletters}
and for the sake of simplicity the electron and pseudofermion
Green's function, and the vertex function were denoted by
$G$, ${\cal G}_M$, and $\Gamma_{MM'}$, respectively.
The initial values for the renormalization factors and couplings are
$Z=Z_M=1$, $K_M=K_M^{(0)}=K M^2$, $J_{MM'}=J_0$ for each $M$, $M'$, at
$D=D_0$.

Using the definition of self energies in Eq.\ (\ref{G}) and (\ref{pG})
the first two equations can be rewritten as
\begin{mathletters}
\begin{equation}
-(\omega-\varepsilon_k-\Sigma_e)\eta+{\partial \Sigma_e\over\partial x}
+\sum\limits_{MM'}\beta_{MM'}
{\partial \Sigma_e\over\partial J_{MM'}}+\sum\limits_M\gamma_M
{\partial \Sigma_e\over\partial K_M}=0
\label{MRGa}\end{equation}
\begin{equation}
-({\tilde\omega}-\lambda-\Sigma_M)\eta_M+{\partial \Sigma_M\over\partial x}+
\sum\limits_{{\tilde M}{\tilde M}'}\beta_{{\tilde M}{\tilde M}'}
{\partial \Sigma_M\over\partial J_{{\tilde M}{\tilde M}'}}+
\sum\limits_{\tilde M}\gamma_{\tilde M}
{\partial \Sigma_M\over\partial K_{\tilde M}}+{dK_M\over dx}=0
\label{MRGb}
\end{equation}
\end{mathletters}
which form is more comfortable for calculating the MRG equations.

\section{Construction of the MRG equations}
\label{sec:3}

To construct the MRG equations the perturbation theory was applied,
that is the Hamiltonian was divided into a non-interacting and an interaction
part with small parameters $J_{MM'}$'s.

For which the electron self energy contains a closed pseudofermion loop,
$\Sigma_e$ tends to zero as $\lambda\rightarrow\infty$.
Thus in the thermodynamical limit for a single impurity from 
Eq.\ (\ref{MRGa}) $\eta=0$ and $Z=1$.

Turning to the other two equations Eq.\ (\ref{MRGb}), (\ref{MRGc})
they were solved in next to leading logarithmic approximation where
the MRG equations are
\begin{equation}
({\tilde\omega}-\lambda)\eta_M^{(2)}=
{\partial\delta\Sigma_M^{(2)}\over\partial x}
+\gamma_M^{(2)}
\label{SE1}\end{equation}
and
\begin{mathletters}
\label{SE2}
\begin{equation}
{\partial \delta\Gamma_{\sigma M,\sigma'M'}^{(2)}\over\partial x}+
\bbox{\sigma}_{\sigma\sigma'}\bbox{S}_{MM'}\beta^{(2)}_{MM'}=0,
\end{equation}
\begin{eqnarray}
({1\over 2}\eta_M^{(2)}+{1\over 2}\eta_{M'}^{(2)}) 
J_{MM'}\bbox{\sigma}_{\sigma\sigma'}
\bbox{S}_{MM'}&+&
{\partial \delta\Gamma_{\sigma M,\sigma'M'}^{(3)}\over\partial x}\nonumber \\
&+&\sum\limits_{{\tilde M}{\tilde M}'}\beta^{(2)}_{{\tilde M}{\tilde M}'}
{\partial \Gamma^{(2)}_{\sigma M,\sigma'M'}\over
\partial J_{{\tilde M}{\tilde M}'}}+
\bbox{\sigma}_{\sigma\sigma'}\bbox{S}_{MM'}\beta^{(3)}_{MM'}=0
\end{eqnarray}
\end{mathletters}
where $\eta_M^{(2)}$, $\gamma_M^{(2)}$ and $\beta^{(2)}_{MM'}$ 
are proportional to the second, $\beta^{(3)}_{MM'}$ to the third 
power of the $J_{MM'}$'s, respectively.
The whole next to leading logarithmic $\beta$-function is $\beta_{MM'}=
\beta^{(2)}_{MM'}+\beta^{(3)}_{MM'}$.

Thus to construct the next to leading logarithmic scaling equations we 
have to calculate the second ($\delta\Gamma_{\sigma M,\sigma'M'}^{(2)}$)
and third ($\delta\Gamma_{\sigma M,\sigma'M'}^{(3)}$)
order vertex corrections, 
and the second order self energy correction ($\delta\Sigma_M^{(2)}$)
for the impurity spin.

These corrections were calculated by applying the thermodynamical Green's
function technique and analytical continuation \cite{AGD}.
Assuming scaling for the vertex function $\Gamma_{MM'}$ 
only one energy variable was kept \cite{Solyom}, thus 
$\omega_1=\omega$, $\omega_2=K_M$, $\omega_3=\omega+K_M-K_{M'}$
and $\omega_4=K_{M'}$, where $\omega_1$, $\omega_2$ 
the incoming, $\omega_3$, $\omega_4$ the outgoing electron and
pseudofermion energies, respectively.

The second and third order vertex diagrams, and the second order correction to
the self-energy for the impurity spin are shown in Fig.\ \ref{fig1}
and Fig.\ \ref{fig3}, respectively.
The detailed calculation of these diagrams is carried out in the Appendix.
Collecting the whole second and third order vertex corrections 
together, they and the self-energy correction for the impurity spin
were substituted into the Eq's\ (\ref{SE1}) and (\ref{SE2}).
In Eq.\ (\ref{SE2}b) the contributions of the third order parquet-type diagrams
depicted in Fig.\ \ref{fig1} (b) and (d) were canceled out with the
terms $\sum\limits_{{\tilde M}{\tilde M}'}\beta^{(2)}_{{\tilde M}{\tilde M}'}
{\partial \Gamma^{(2)}_{\sigma M,\sigma'M'}\over
\partial J_{{\tilde M}{\tilde M}'}}$ as the leading
logarithmic scaling equations are equivalent to the summing up of the
parquet diagrams.
The divergences at finite $T$ were canceled out, too.
Thus only the diagram Fig.\ \ref{fig1} (c) contributes to Eq.\ (\ref{SE2}b).

Introducing the dimensionless couplings $j_{MM'}=\rho_0 J_{MM'}$
the next to leading logarithmic scaling equations are
\begin{eqnarray}
\eta_M={d\ln Z_M\over dx}&=&q^2(S,M) j_{M,M-1} j_{M-1,M}\Theta_{M,M-1}(D)
\nonumber \\ &+&
p^2(S,M) j_{M,M+1} j_{M+1,M}\Theta_{M,M+1}(D)+2 M^2 j^2_{M,M}
\label{ujSE1}\end{eqnarray}
\begin{eqnarray}
\gamma_M={dK_M\over dx}&=&(K_{M-1}-K_M) q^2(S,M) 
j_{M,M-1} j_{M-1,M}\Theta_{M,M-1}(D)
\nonumber \\ &+&
(K_{M+1}-K_M) p^2(S,M) j_{M,M+1} j_{M+1,M}\Theta_{M,M+1}(D)
\label{ujSE2}\end{eqnarray}
\begin{eqnarray}
&&\rho_0\beta_{M,M+1}={dj_{M,M+1}\over dx}=-(M j_{M,M}-(M+1) j_{M+1,M+1})
j_{M,M+1}\bigl(1+\Theta_{M,M+1}(D)\bigr)\nonumber \\
&+&q^2(S,M) j_{M,M-1} j_{M-1,M} j_{M,M+1}\Theta_{M,M-1}(D)
\biggl (\Theta_{M,M+1}(D)-{1\over 2}\biggr )
\nonumber \\
&+&p^2(S,M+1) j_{M,M+1} j_{M+1,M+2} j_{M+2,M+1}\Theta_{M+1,M+2}(D)
\biggl (
\Theta_{M,M+1}(D)-{1\over 2}\biggr )
\nonumber \\
&-&p^2(S,M) j^2_{M,M+1} j_{M+1,M}\Theta_{M,M+1}(D)
-M^2 j^2_{M,M} j_{M,M+1}\nonumber \\
&-&(M+1)^2 j^2_{M+1,M+1} j_{M,M+1}+2 M (M+1) j_{M,M} j_{M,M+1} j_{M+1,M+1}
\label{ujSE3a}\end{eqnarray}
\begin{eqnarray}
&&\rho_0\beta_{M,M-1}={dj_{M,M-1}\over dx}=-((M-1) j_{M-1,M-1}-M j_{M,M})
j_{M,M-1}\bigl(1+\Theta_{M,M-1}(D)\bigr)\nonumber \\
&+&p^2(S,M) j_{M,M+1} j_{M+1,M} j_{M,M-1}\Theta_{M,M+1}(D)\biggl (
\Theta_{M,M-1}(D)-{1\over 2}\biggr )
\nonumber \\
&+&q^2(S,M-1) j_{M,M-1} j_{M-1,M-2} j_{M-2,M-1}\Theta_{M-1,M-2}(D)\biggl (
\Theta_{M,M-1}(D)-{1\over 2}\biggr )
\nonumber \\
&-&q^2(S,M) j^2_{M,M-1} j_{M-1,M}\Theta_{M,M-1}(D)
-M^2 j^2_{M,M} j_{M,M-1}\nonumber \\
&-&(M-1)^2 j^2_{M-1,M-1} j_{M,M-1}+2 M (M-1) j_{M,M} j_{M,M-1} j_{M-1,M-1}
\label{ujSE3b}\end{eqnarray}
and for $M\neq 0$
\begin{eqnarray}
&&\rho_0\beta_{M,M}={dj_{M,M}\over dx}\nonumber \\
&=&{1\over M}
\biggl [
q^2(S,M) \Theta_{M,M-1}(D) j_{M,M-1} j_{M-1,M}
-p^2(S,M) \Theta_{M,M+1}(D) j_{M,M+1} j_{M+1,M}
\biggr ]
\nonumber \\
&+&q^2(S,M){M-1\over M} j_{M,M-1} j_{M-1,M-1} j_{M-1,M}\Theta_{M,M-1}(D)
\nonumber \\
&+&p^2(S,M){M+1\over M} j_{M,M+1} j_{M+1,M+1} j_{M+1,M}\Theta_{M,M+1}(D)
\nonumber \\
&-&q^2(S,M) j_{M,M-1} j_{M-1,M} j_{M,M}\Theta_{M,M-1}(D)\nonumber \\
&-&p^2(S,M) j_{M,M+1} j_{M+1,M} j_{M,M}\Theta_{M,M+1}(D)
\label{ujSE3c}\end{eqnarray}
where $\Theta_{MM'}(D)$ ensures that $D>\sqrt{(K_M-K_{M'})^2+T^2}$
with the definition
\begin{equation}
\Theta_{MM'}(D)=\cases{1&if $D>\sqrt{(K_M-K_{M'})^2+T^2}$\cr
0&if $D<\sqrt{(K_M-K_{M'})^2+T^2}.$\cr}
\end{equation}
The definition of $p(S,M)$ and $q(S,M)$ are given in Eq.\ (\ref{Spm}). 
It must be stressed that these scaling equations are valid for
$T\leq D$.

\section{Solution of the scaling equations}
\label{sec:4}

It can be seen from Eq.\ (\ref{ujSE2}), that the corrections to the bare
$K_M^{(0)}$ are proportional to the second power of $j_{MM'}$'s in the 
leading order.
In the other equations $K_M$ appears only in the arguments of the
$\Theta$ functions which are multiplied by the second or third power
of $J_{MM'}$'s. Thus to keep in the approximation consistent
$K_M=K_M^{(0)}=K M^2$ can be taken in the arguments of the $\Theta$ functions.
After this consideration the solution of the equations becomes simpler, 
because the equations for $\beta_{MM'}$ are not coupled.
Exploiting the symmetries of the scaling equations
\begin{mathletters}
\begin{equation}
j_{M,M'}=j_{M',M}
\end{equation}
\begin{equation}
j_{M,M'}=j_{-M,-M'}
\end{equation}
\end{mathletters}
must hold.

Thus the equations which have to be solved are
\begin{eqnarray}
{dj_{M,M+1}\over dx}&=&-(M j_{M,M}-(M+1) j_{M+1,M+1})
j_{M,M+1}\bigl(1+\Theta_{M,M+1}(D)\bigr)\nonumber \\
&+&q^2(S,M) j^2_{M,M-1} j_{M,M+1}\Theta_{M,M-1}(D)
\biggl (\Theta_{M,M+1}(D)-{1\over 2}\biggr )
\nonumber \\
&+&p^2(S,M+1) j_{M,M+1} j^2_{M+1,M+2} \Theta_{M+1,M+2}(D)
\biggl (\Theta_{M,M+1}(D)-{1\over 2}\biggr )
\nonumber \\
&-&p^2(S,M) j^3_{M,M+1} \Theta_{M,M+1}(D)
-M^2 j^2_{M,M} j_{M,M+1}\nonumber \\
&-&(M+1)^2 j^2_{M+1,M+1} j_{M,M+1}+2 M (M+1) j_{M,M} j_{M,M+1} j_{M+1,M+1}
\label{E1}\end{eqnarray}
and for $M\neq 0$
\begin{eqnarray}
{dj_{M,M}\over dx}&=&{1\over M}\biggl [
q^2(S,M) \Theta_{M,M-1}(D) j^2_{M,M-1}\nonumber \\
&-&p^2(S,M) \Theta_{M,M+1}(D) j^2_{M,M+1}
\biggr ]
\nonumber \\
&+&q^2(S,M){M-1\over M} j^2_{M,M-1} j_{M-1,M-1} \Theta_{M,M-1}(D)
\nonumber \\
&+&p^2(S,M){M+1\over M} j^2_{M,M+1} j_{M+1,M+1} \Theta_{M,M+1}(D)
\nonumber \\
&-&q^2(S,M) j^2_{M,M-1} j_{M,M}\Theta_{M,M-1}(D)
\nonumber \\
&-&p^2(S,M) j^2_{M,M+1} j_{M,M}\Theta_{M,M+1}(D)
\label{E2}\end{eqnarray}

These equations were solved numerically for different initial couplings
$j_0=j_{MM'}(D=D_0)$. The results for $j_0=0.1$ and $j_0=0.0435$ at $K/T=10$ 
are shown in Fig.\ \ref{fig4} and Fig.\ \ref{fig5} for $S=2$ and $S=5/2$, 
respectively. The initial bandwidth cutoff was chosen as $D_0=10^5$\,K. 

\section{Resistivity}
\label{sec:5}

The Kondo resistivity was calculated by solving the Boltzmann equation
in the presence of the spin-orbit induced anisotropy, using the value of
running couplings ($j_{MM'}$) calculated in the preceding section, 
at $D=T$.

Taking the usual form \cite{Yosida} for the electron distribution function 
$f(\varepsilon_k)$ in the presence of the electric field ${\bf E}$ as
\begin{equation}
f(\varepsilon_k)=f_0(\varepsilon_k)-({\bf k}{\bf E})
\Phi(\varepsilon_k) {\partial
f_0(\varepsilon_k)\over\partial\varepsilon_k}, 
\label{distrib}\end{equation}
the Kondo contribution to the resistivity is
\begin{equation}
{1\over\rho_{\text{Kondo}}}=-{e\over 3\pi^2}(2 m)^{3/2}
\int d\varepsilon_k \varepsilon_k^{3/2} \Phi(\varepsilon_k)
\biggl(-{\partial
f_0\over\partial\varepsilon_k}\biggr)
\label{cond}\end{equation}
where the function $\Phi$ is determined by the Boltzmann equation
\begin{equation}
{e\over m}{\partial f_0(\varepsilon_k)\over\partial\varepsilon_k}
({\bf k}{\bf E})
+\biggl({\partial f\over\partial t}\biggr)_{\text{coll.}}=0.
\label{BE}\end{equation}
The collision term $\bigl({\partial f\over\partial t}\bigr)_{\text{coll.}}$
can be expressed in terms of transition probabilities as
\begin{eqnarray}
\biggl({\partial f\over\partial t}\biggr)_{\text{coll.}}&=&
{c\over V}\sum\limits_{{\bf k'}\sigma'}\bigl\{W({\bf
k'},\sigma'\rightarrow {\bf k},\sigma) f({\bf
k'})(1-f({\bf k}))\nonumber \\
&-&W({\bf k},\sigma\rightarrow {\bf k'},\sigma') f({\bf
k})(1-f({\bf k'}))\bigr\}
\label{coll}\end{eqnarray}
where e.g. $W({\bf k'},\sigma'\rightarrow {\bf k},\sigma)$ represents 
the transition probability from a ${\bf k'},\sigma'$ state  to a 
${\bf k},\sigma$, $c$ is the impurity concentration, and $V$ is the volume. 

Turning to our case, these probabilities can be calculated as
\begin{equation}
W({\bf k},\sigma\rightarrow {\bf k'},\sigma')=\sum\limits_{MM'}
p_M w({\bf k},\sigma,M\rightarrow {\bf k'},\sigma',M')
\end{equation}
where
\begin{equation}
w({\bf k},\sigma,M\rightarrow {\bf k'},\sigma',M')=2\pi
\bigl|T_{{\bf k},\sigma,M\rightarrow {\bf k'},\sigma',M'}\bigr|^2
\delta(\varepsilon_k-\varepsilon_{k'}+K M^2-K M'^2),
\label{kisw}\end{equation}
and $p_M=e^{-\beta K M^2}/Z$, $\beta=1/T$, 
$Z=\sum\limits_M e^{-\beta K M^2}$. 

The scattering amplitude in Eq.\
(\ref{kisw}) is expressed in terms of the renormalized couplings 
$J_{MM'}(x)$ as 
\begin{equation}
T_{{\bf k},\sigma,M\rightarrow {\bf
k'},\sigma',M'}\approx J_{MM'}(x=\ln{D_0\over
T})\bbox{\sigma}_{\sigma\sigma'} \bbox{S}_{MM'}
\end{equation}
where the dependence on the direction of the momenta ${\bf k}$ and
${\bf k'}$ is ignored and that makes the Boltzmann equation solvable 
in a simple form. The k-dependence may result as some numerical 
factors in the final expression, but in the main features of the
temperature dependence those are not playing an important role.

Substituting these assumptions into Eq.\ (\ref{coll}), 
changing the sum ${1\over V}\sum\limits_{k'}$ to $\int
\rho(\varepsilon_{k'}) d\varepsilon_{k'}\int {d\Omega_{k'}\over
4\pi}\approx$ $\rho_0\int\limits_{-D_0}^{D_0} d\varepsilon\int
{d\Omega_{k'}\over 4\pi}$, using the properties of the spin algebra for
$\bbox{\sigma}$ and $\bbox{S}$ and the "detailed balance" principle
($\bigl({\partial f_0(\varepsilon_k)\over\partial t}\bigr
)_{\text{coll.}}=0$), we obtain after linearization in ${\bf E}$
\begin{equation}
\biggl({\partial f\over\partial t}\biggr)_{\text{coll.}}=
{2\pi c\over \rho_0}({\bf k}{\bf E})\Phi(\varepsilon_k)
{\partial f_0\over\partial\varepsilon_k} F(\varepsilon_k)
\label{coll1}\end{equation}
where
\begin{eqnarray}
F(\varepsilon_k)&=&\sum\limits_M p_M\biggl\{M^2 j^2_{M,M}(x=\ln{D_0\over T})
\nonumber \\
&+&(S(S+1)-M(M+1)) j^2_{M,M+1}(x=\ln{D_0\over T})\nonumber \\
&&\cdot\bigl[1-(1-e^{-(2 M+1) K/T}) f_0(\varepsilon_k-(2 M+1) K)\bigr]\biggr\}
\label{F}\end{eqnarray}
where we introduced the dimensionless coupling constants $j_{MM'}=\rho_0
J_{MM'}$.
Inserting Eq.\ (\ref{coll1}) into the Boltzmann Eq.\ (\ref{BE})
we obtain for $\Phi$
\begin{equation}
\Phi(\varepsilon_k)=-{e\over m}\biggl [{2\pi c\over\rho_0}
F(\varepsilon_k)\biggr ]^{-1}
\end{equation}
and for the Kondo resistivity
\begin{equation}
{1\over\rho_{\text{Kondo}}}={1\over\rho^{(0)}}
\int d\varepsilon \biggl(-{\partial
f_0\over\partial\varepsilon}\biggr) F^{-1}(\varepsilon)
\label{cond1}\end{equation}
where the usual assumption 
\begin{equation}
\int d\varepsilon\varepsilon^{3/2}
\biggl(-{\partial f_0\over\partial\varepsilon}\biggr)
F^{-1}(\varepsilon)\approx \varepsilon_F^{3/2} \int d\varepsilon 
\biggl(-{\partial f_0\over\partial\varepsilon}\biggr) F^{-1}(\varepsilon)
\end{equation}
was taken into account and the constant $\rho^{(0)}
={3\over 4}{m\over e^2}{2\pi c\over\varepsilon_F\rho_0^2}$ was introduced.

In the case of $K=0$, Eq.\ (\ref{cond1}) reproduces the bulk
Kondo resistivity.

The resistivity was calculated by evaluating the occuring integral
in Eq.\ (\ref{cond1}) numerically for different $K$ values which are in 
the regime discussed in
Part I \cite{Ujsaghy1}. The resistivity as the function of the
temperature is shown in Fig.\ \ref{fig6} (a) and (b) for $S=2$
and $S=5/2$, respectively. The plots are similar to the 
experimental ones (see Sec.~\ref{sec:7}). 

The effect of the anisotropy on the Kondo temperature defined by the 
largest slope in the resistivity can be examined by looking at the 
derivative of the calculated resistivity vs. temperature. We can see
from Fig.\ \ref{fig6} (a) and (b) that the Kondo temperature defined 
in that way is only slightly affected by the anisotropy in those cases 
where the Kondo effect is pronounced (e.g. $K<0.5$ K in Fig.\ \ref{fig6}).
The effect of the anisotropy becomes dominant for larger strength of $K$.
The temperature dependence of the resistivity has a maxima depending on 
the strength of $K$ and the spin-flip contribution is freezing out gradually.
That behavior is very different for integer and half-integer spins
for large anisotropy. For integer spins the impurity contribution tends
to zero in a way which is very sensitive on the strength of the anisotropy. In
the case of half-integer spin the impurity resistivity approaching, however, 
a finite value at zero temperature which is independent of $K$. The resistivity 
there is determined by the dynamics of the two lowest energy levels shown in
Fig.\ \ref{fig02}. In those cases the Kondo effect is also essentially reduced 
due to the smallness of the spin-flip amplitudes, but still presents.
(The Kondo contribution in the lowest order is proportional to 
$\bbox{S}\bbox{\sigma}$ which gives a small amplitude for $M=\pm 1/2$.)

It is important to emphasize, that in case $S=1/2$ the anisotropy is loosing 
its meaning. 

In a real system the anisotropy strength $K$ has a distribution thus the 
formation of the resistivity maximum at finite value of temperature cannot 
be expected at least above the Kondo temperature. The calculation is anyhow 
not reliable for $T<T_K$.

\section{Kondo resistivity in thin films}
\label{sec:6}

To get some information about the case of thin films a simple
assumption is made that the two surfaces contribute to the
anisotropy constant $K$ in an additive way.
The anisotropy factor for a sample with thickness $t$ and in a
distance $d$ measured from one of the surface is
\begin{equation}
K(d,t)=K_d+K_{t-d}={\alpha\over d}+{\alpha\over t-d}
\label{kate}\end{equation}
where the coefficient $\alpha$ is estimated in Part I \cite{Ujsaghy1}
(see Eq.\ (32)).
The appropriate calculation of the resistivity including
the elastic impurity scattering with mean free path $l_{\text{el}}$
is a very difficult task for a film for an
arbitrary ratio of $l_{\text{el}}/t$ and value of $K$.
In order to avoid those difficulties we are
making use of the fact, that the magnetic exchange (Kondo)
contribution to the resistivity $\rho_{\text{Kondo}}$ is smaller
by a factor $10^{-3}$ than the residual normal impurity
resistivity $\rho_{\text{\text{nor}}}$
($\rho=\rho_{\text{\text{nor}}}+\rho_{\text{Kondo}}$),
thus an expansion in the Kondo contribution is appropriate. The
calculation can be carried out in two limits (i) $t<l_{\text{el}}$ and
(ii) $t\gg l_{\text{el}}$. It will be shown that the final expression
does not depend on which limit is considered. In the case (i)
the electrical resistivity contains the average value of the
inverse electron life time. Denoting the resistivity at
temperature $T$ for a given value of $K$ by $\rho(K,T)$, the
average over the value of $K(d,t)$ is
\begin{equation}
{\bar\rho}(t,T)={1\over t}\int\limits_0^t\rho(K(x,t),T) dx.
\label{roatlag}\end{equation}
On the other hand, in case (ii) the sample can be considered as
a set of parallel resistors of equal size, where each resistor represents a
stripe in the sample with a constant $K$. In that case the
conductances are additive, thus
\begin{equation}
{\bar\rho}(t,T)={1\over \sigma_{\text{nor}}+{1\over
N}\sum\limits_i\sigma_i(K(x_i,t),T)} 
\end{equation}
where $N$ is the number of the resistors (stripes) labeled by
$i$ and $\sigma_i$ represents the Kondo conductivity of stripe $i$
placed in distance $x_i$. In the actual case only the first stripes 
depend on the surface anisotropy. The Kondo conductivity is defined by
the Kondo resistivity given by Eq.\ (\ref{cond1}) as 
\begin{equation}
\sigma=\sigma_{\text{nor}}+\sigma_{\text{Kondo}}=(\rho_{\text{nor}}+
\rho_{\text{Kondo}})^{-1}
\end{equation}
where
$\sigma_{\text{Kondo}}\approx-{\rho_{\text{Kondo}}\over\rho_{\text{nor}}^2}<0$.
The expansion gives the final expression
\begin{equation}
{\bar\rho}(t,T)=\rho_{\text{nor}}+{1\over
t}\int\limits_0^t\rho_{\text{Kondo}}(K(x,t),T) dx.
\label{ugyanaz}\end{equation}
That expression valid in the limit
$\rho_{\text{Kondo}}\ll\rho_{\text{nor}}$ 
gives back exactly the expression in Eq.\ (\ref{roatlag}).

In the numerical calculation the integral in Eq.\
(\ref{roatlag}) or (\ref{ugyanaz}) is replaced by a
weighted sum with appropriate intervals.
Introducing the new integration variable $d/\alpha$,
the calculated Kondo resistivity depends only on $t/\alpha$
which is shown in Fig.\ \ref{figatlagolt} (a) and (b) for $S=2$ and $S=5/2$,
respectively.
Fitting the calculated Kondo resistivity for temperatures $T\gg T_K$ 
($T=2-4$ K) by the function 
$\rho_{\text{Kondo}}/\rho^{(0)}=-B_{\text{calc}}\ln T$ 
as it has been done in
the experimental works (see Sec.~\ref{sec:7}), 
the behavior of $B_{\text{calc}}$  was examined as a function of 
$t/\alpha$ which can be seen in Fig.\ \ref{fig10}. 
To compare this calculated dependence of the coefficient $B$ on the thickness
to the experimental data, they were fitted by the function 
$B(t)=\rho^{(0)} B_{\text{calc}}(t/\alpha)$ as it is shown in 
Fig.\ \ref{kisfit}.
The fitted value of $\alpha$ is $\alpha=247.7\text{\AA}$ K which
is in agreement with the prediction given in Part I 
\cite{Ujsaghy1} by Eq.\ (32) (see Sec.~\ref{sec:7}). The fit is not 
too sensitive to small changes ($<5\%$) in $\alpha$.

If the sample is not thin, then the above results can be phenomenologically 
described in the framework of a simple model where the impurities in the
region of the surface do not contribute to the Kondo resistivity and
outside that region they are not affected. In this way the effective 
suppression length $\lambda$ can be introduced and then the average
resistivity at low temperature $\rho_{t}$ e.g. for a thickness
$t$
\begin{equation}
\rho_t=\rho_{t=\infty}{t-2\lambda\over t}.
\label{efflambda}
\end{equation}

According to this semi-phenomenological formula $B(t)=B_{\infty} (1-2\lambda/t)$ 
which was fitted to the experimental data. This can be seen also in 
Fig.~\ref{kisfit} where the fitted value of the effective suppression layer 
parameter is $\lambda=207.5\text{\AA}$. 

The effect of the mean free path in the ballistic region can be demonstrated 
directly by taking into account the effect of the mean free path in
the anisotropy constant. We calculated the change of the electrical resistivity for a thin film with thickness $L=600\text{\AA}$ with anisotropy arising
only at one of the surfaces in the forms
\begin{mathletters}
\label{apt}
\begin{equation}
K=A\ {\xi\over d}
\end{equation}
\begin{equation}
K=A\ {\xi\over d}\ e^{-d/\xi}
\end{equation}
\end{mathletters}
where $\xi$ is the elastic mean free path (e.g. $\xi=100\text{\AA}$),
$d$ is the distance measured from the surface with anisotropy of strength $A$,
and the exponential decay is due to the mean free path. The electrical
resistivity is calculated for $S=2$ at $T=2 T_K$ just above 
the Kondo temperature $T_K$,
as a function of the strength $A$ of the anisotropy for two cases without and 
with exponential factor (see Eq.\ (\ref{apt})).
Increasing the anisotropy strength $A$ the spins are completely frozen in 
nearby the surface, but that region is limited by the finite mean free path.
Fig.\ \ref{aperte} clearly demonstrates that the strength of the
anisotropy and the size of the suppression layer are reduced due to
the finite mean free path as it is calculated by taking into account
the anisotropy only for one of the surfaces.

\section{Comparison with experiments}
\label{sec:7}

In the last couple of years a very extensive study of the Kondo 
effect in thin films and wires have been performed.
The experimental works were concentrating on determination of the
effect of reduced dimensions on the Kondo temperature $T_K$ and the
amplitude of the resistivity anomaly. A detailed critical discussion
of the earlier works are given in \cite{Blachy}. The early studies
have been performed by Giordano and his collaborators \cite{Chen1,Chen2},
and by DiTusa, Lin, Park, Isaacson and Parpia \cite{DiTusa}.
In order to discuss the effect of uncoupled magnetic impurities,
first only those experiments are listed which are performed in
dilute limit, thus
e.g. for Au(Fe) alloys the Fe concentration is $30$ ppm.
These experiments belong to two groups depending whether size effect 
was observed or not.

Concerning the theory two regions must be distinguished. When the
size of the sample (e.g. the thickness of the sample) is inside the 
ballistic region, then obviously the present theory must be applied.
In the case of thicker samples more care must be paid.
There is another theory by Martin, Wan and Phillips \cite{Martin}
which is applicable
in the opposite limit of weak localization, where the disorder-induced
depression or enhancement of the Kondo effect is predicted depending
on the value of the spin flip scattering rate $\tau_s^{-1}$
(depression is the case where $T_K$, $\hbar\tau_s^{-1}\ll T$).
The competition between these theories needs further studies.

In the following the discussion is organized according to different effects.
First we discuss how the change in the density of states at the surface
can influence the Kondo effect, but it is ruled out as an explanation of 
the size effects to be discussed, because it is applicable only on much 
smaller scale (Sec.~\ref{sub:A}). Then the experiments with considerable
dependence on the size of the samples are discussed and compared with the 
present theory (Sec.~\ref{sub:B}). Finally those experiments are listed where 
no size effect was observed (Sec.~\ref{sub:C}) or the concentrations of 
the impurities are in the spin glass region (Sec.~\ref{sub:D}).

\subsection{Density of states effects}
\label{sub:A}

As it has been discussed in Sec.~\ref{sec:1} of Part I \cite{Ujsaghy1}, 
the size dependence cannot be
expected just because the Kondo cloud cannot fully develop in all directions
by reducing the size of the sample. The only possibility which has
been discussed by Zar\'and \cite{Zarand} is, that nearby the surface there
is a change in the density of states of conduction electrons by formation
of a Friedel type oscillation due to the surface. That explanation
was ruled out, because those changes in the density of states are very
much localized in a few atomic distances measured from the surface
and the smallest sizes in the experiments to be discussed are about 
$300\text{\AA}$. That effect may, however, show up in point contact 
experiments where the contact size is smaller by even more than one 
order of magnitude. Such experiments were performed by Yanson and
his collaborators \cite{Yanson1,Yanson2,Yanson3} with Mn and Fe
impurities in Cu contacts. Zar\'and and Udvardi \cite{Udvardi1,Udvardi2}
showed that depending 
on the actual position of the impurity the density of states for an
essential energy range around the Fermi surface can be enhanced or
depressed by even $20\%$, thus $\rho=\rho_0+\delta\rho$, 
where $|\delta\rho/\rho_0|<0.2$. 
Just in order to demonstrate the effect an energy independent $\delta\rho$
is assumed and for that case 
in the expression of the Kondo temperature $\exp[-1/{2 J(\rho_0+
\delta\rho)}]=\exp[-1/{2 J\rho_0}] \exp[(2 J\rho_0)^{-1}{\delta\rho
\over\rho_0}]$ there is an enhancement due to the second factor. 
Depending on the value of
$(J\rho_0)^{-1}$ that enhancement can be over a factor of $100$
for Mn and about $2-3$ for Fe impurities. The enhancement is the larger the
smaller the Kondo temperature $T_K$ \cite{Yanson1,Yanson2,Yanson3,Udvardi2}.
In the experiments the enhancement is the
larger the smaller the contact size, thus to have large enhancement
most of the impurities must be nearby the surface. 
Similar effect was also seen \cite{Keijsers} in point contacts with
presumable tunneling two-level systems (TLS) where an atom jumps 
between two positions and the orbital Kondo effect is developed
\cite{Zawa1,Zawa2} by coupling the conduction electrons with different
angular momenta to the TLS. As the typical size of the studied films and
wires are much larger and such a dominating enhancement of the Kondo
temperature has never been observed, therefore that explanation
can be ruled out.

\subsection{Experiments with observed size effect}
\label{sub:B}

Giordano and his collaborators (see for a review \cite{Blachy}) have
performed a series of different experiments under different
conditions where the size effect was observed but the changes in the
Kondo temperature were almost negligible. The experiments of different type
are listed below.
\begin{description}
\item[(i)] Dependence on film thickness. 

The film experiments with thickness $265\text{\AA}-1800\text{\AA}$
were performed e.g. with $30$ ppm Fe in Au, but similar results are obtained 
also for $100$ ppm \cite{Chen1,Chen2}. The resistivity was fitted by the
formula
\begin{equation}
\rho(T)=\rho-B\ln(T)
\end{equation}
where $B$ is an adjustable parameter. It is well known for the Kondo 
systems that $B$ is just not the result of the first non-vanishing
third order perturbational result where $B$ would be $B\sim J^3$,
but it is the actual slope nearby or somewhat above the Kondo temperature
(see for example \cite{Blachy}). In the actual experiments the 
temperature range $1.8-4$ K
was studied while $T_K=0.3$ K. The dependence of that coefficient $B$ on
thickness was plotted as shown in Fig.\ \ref{kisfit}.
The experimental results are fitted by the calculated dependence of $B$ on
the thickness with parameter $\alpha=247.7\text{\AA}$ K and 
by the semi-phenomenological formula 
given by Eq.\ (\ref{efflambda}) with the effective suppression layer
parameter value $\lambda=207.5\text{\AA}$ in Fig.\ \ref{kisfit}
\cite{Blachy1,Blachy2}. That value of $\alpha$ is in agreement with
the estimate given in Part I \cite{Ujsaghy1} by Eq.\ (32).
There was not any signal for essential change in the Kondo temperature
\cite{Blachy} in agreement with our theoretical result. 
It is interesting to note that the estimated 
Kondo coherence length was about $3\cdot 10^4\text{\AA}$, much larger than the 
thickness of the sample.
Similar experiments were performed with wires where more geometrical
effects are expected, but the results are qualitatively similar but 
not identical.
The simple semi-phenomenological formula given by Eq.\ (\ref{efflambda}) is 
not appropriate in those cases.
Qualitatively very similar results are reported in \cite{Apostol}
but there are quantitative differences very likely due to the sample
preparation.
\item[(ii)] Kondo proximity effect.

A set of experiments \cite{Blachy3,Blachy4} were performed where
the film of dilute alloys is covered by a second layer of pure metal.
The observation was, that in the case of a thin layer of dilute alloys
with a significant suppression of the Kondo effect the covering by a
second pure film results in partial recovery of that suppression.
In Fig.\ \ref{fig11} with the suppression layers indicated it is shown that
the bilayer structure has a suppression layer only on one side of the 
film of dilute alloys, thus only one half of the suppression is expected.
In order to verify the importance of the role of the spin-orbit interaction in
the superimposed layer to complete
the neighborhood of the impurity with a uniform spin-orbit coupling,
we suggest experiments where the superimposed layer has negligible spin-orbit
interaction (e.g. Al or Mg). In that case the boundary is changed, but the
anisotropy should remain.
\item[(iii)] Kondo proximity effect with overlayers with different
disorder.

It has been shown experimentally \cite{Blachy5} that the Kondo resistivity
suppression in a film of dilute alloys covered by a pure film but with
different disorder depends on the disorder in the overlayer.
It was found that the larger the disorder
the smaller the recovery is.
As it is discussed above if the thickness of the overlayer and the mean 
free path
$l_{\text{el}}$ in it are larger than the thickness of the suppression layer
$\lambda$, then the depression takes place only on one side of the
film of dilute alloys.
On the other hand, if the pure overlayers contains disorder, then the
electron entering that overlayer cannot bring back information to the
magnetic impurity by their momenta, as their momenta is changed
in the overlayer (the overlayer is not in the ballistic regime).
In these cases the reduction in the anisotropy is only partially 
developed as the surroundings of the impurity is not perfectly spherical
in contrast to the case of overlayer with long mean free path.
\end{description}

\subsection{Experiments without size effect}
\label{sub:C}

In contrast to the measurements discussed in \ref{sub:B} there is a 
series of experiments by 
Chandrasekhar {\it et al.} \cite{Chandra} where the size dependence 
was not found. The geometries of these experiments were different, the
thickness of the sample $t$ was kept the same ($t=380\text{\AA}$),
but the width of the stripes $w$ was changed between $380-10^6\text{\AA}$
(see Fig.\ \ref{fig12}).
After the correction due to the weak localization effects and due to 
electron-electron interaction no size dependence was claimed. On
the basis of the present theory, for samples
$t\ll w$ no size dependence is expected as the ratio of the volume
of the suppression layers to the total volume is not changing. 
Where $t\sim w$ the anisotropy due to the geometry becomes more
complicated, thus it is hard to make comparison with the present
theory. On the other hand for $t\sim w$ the experimental points
are somewhat falling off from the main averaged line, that, of course, may
be due to experimental errors. According to the present theory the 
averaged Kondo resistivity for $w\gg t$ had to be smaller than the
bulk resistivity, 
but it seems to be not the case \cite{Chandra2}.

Finally it should be mentioned that no size effect was observed studying
${\rm La_{\text{1-x}}Ce_{\text{x}}}$ films where Ce has $S=1/2$ for which 
no surface anisotropy is expected \cite{Roth}.

\subsection{Higher concentration}
\label{sub:D}

There are several experiments \cite{DiTusa,Neuttiens} with higher
impurity concentration. In these cases the impurity-impurity
interaction mediated by the RKKY interaction competes with the
Kondo effect. In another set of experiments \cite{DiTusa} the thickness of
the film were changed in samples made of Cu with $1000$ ppm Cr and
it was found very similar depression of the Kondo effect described above 
in Sec.~\ref{sub:B}. The wires with geometries similar to those 
discussed in Sec.~\ref{sub:C} but with $2800$ ppm impurities do not show
dependence on the width $d$ \cite{Neuttiens}, but the overall
amplitude is substantially SU-pressed compared to the bulk, which was
attributed to spin-glass effects.

\section{Conclusion}
\label{sec:8}

In the present paper the influence of the spin-orbit induced surface anisotropy
is studied on the Kondo effect in dilute magnetic alloys samples of 
finite size at least in one dimension.
That anisotropy splits the energy levels for impurity spin $S>1/2$. That
anisotropy reduces as the bulk part of the sample is approached
relatively slowly as $1/d$ where $d$ is the distance of the impurity
measured from the surface. That anisotropy occurs for samples of any shape,
but for those cases further theories should be developed. The range where 
the anisotropy is relevant can be characterized by the suppression length 
$\lambda$ introduced in Sec.~\ref{sec:6}, which is proportional to the strength
of the anisotropy but limited by the mean free path of the electron as
the anisotropy reflects the presence of the surface in the ballistic region 
nearby the impurity. Thus that suppression length cannot exceed a few hundred 
$\text{\AA}$ in accordance with the experiments discussed in Sec.~\ref{sub:B}.

That anisotropy hinders the motion of the impurity spin $S$ if $S>1/2$
and the Kondo effect is affected in those regions of the samples where the
anisotropy is not negligible relative to the Kondo temperature $T_K$.
In order to calculate the Kondo resistivity the renormalized exchange
coupling constants are calculated in Sec.~\ref{sec:3} and \ref{sec:4}
by using the multiplicative renormalization group technique which is
applicable only for temperatures $T$ larger than the Kondo temperature $T_K$,
thus no detailed prediction can be made outside that region.
It can be accepted, however, that if the Kondo effect is already reduced
in region $T>T_K$ similar effect is expected also for $T<T_K$.
The resistivity is calculated by solving the Boltzmann equation
in Sec.~\ref{sec:5} for integer and half-integer spins
with different anisotropy strength. Even if the calculated resistivity
curves in Fig.~\ref{fig6} (a) and (b) show different characteristic
features by developing a resistivity maxima at different temperatures
and of different amplitudes, those features are almost loosen as an average 
over the strength of the anisotropy is taken for $T>T_K$. The curves 
calculated for 
thin films (see Fig.~\ref{figatlagolt}) show smooth increase of the resistivity.
More structures could be expected only in those experiments where the 
impurities are in certain distance measured from the surface.
If the anisotropy does not dominate the complete sample then as the result 
of the average taken the largest resistivity slope as a function of temperature
is in the region of the Kondo temperature $T_K$, and its position
cannot be shifted too much on the scale of the Kondo temperature $T_K$. That 
theoretical result is in accordance with the experimental findings
(see Sec.~\ref{sub:B}).

The relatively weak sensitivity of the observed region of the largest 
resistivity slope on the size of the samples rules out the density states
effects nearby the surface in contrast to the point contact experiments 
(see Sec.~\ref{sub:A}). The size dependence associated with the large Kondo
compensation cloud is not observed in agreement with the Kondo theory where
such a simple connection is ruled out.

The calculated Kondo resistivity for thin films was fitted for temperatures
$T\gg T_K$ ($T=2-4$ K) by the function
$\rho_{\text{Kondo}}/\rho^{(0)}=-B_{\text{calc}}\ln T$
which is compared to the experimental data in Fig.~\ref{kisfit} and
gives excellent agreement.

The phenomenological theory using the effective suppression length $\lambda$
(see Sec.~\ref{sec:6}) works remarkable well to interpret qualitatively 
the experimental
data quoted in Sec.~\ref{sec:7}. The fit of the experimental data is shown
in Fig.~\ref{kisfit}.

The different proximity effects described in Sec.~\ref{sub:B} can be also
well explained by the present theory.

It is important that the role of mean free path (Sec.~\ref{sec:6},
see Fig.~\ref{aperte}) reduces the effect of the large anisotropy
constant, thus for a large range of strong anisotropy the size dependence
remains in a limited range as far as the elastic mean free paths are
in the same order of magnitude. In this way the size effect can be
comparable for different host materials with different large spin-orbit 
interactions but with comparable elastic mean free path.

We have to emphasize, however, that our calculation does not consider the
localization effects which are present in samples of larger sizes.
Such effect have recently been predicted by Martin, Wan and Phillips
\cite{Martin} and deserves further detailed studies.

In addition to those localization effects the theoretical studies
must be extended to the microscopic calculations of the anisotropy
constant.

Considering further experiments the mean free path effects should be studied.
The most relevant experiment to directly verify the role of the spin-orbit
interaction could be the proximity experiments where the superimposed layer 
would be made of another metal without spin-orbit interaction as it is 
discussed in Sec.~\ref{sub:B}. In those cases the uniform surrounding of 
the impurity would not be developed, thus the anisotropy remains. Furthermore, 
the experiments with impurities in a certain distance measured from the
surface would be also very instructive.

Summarizing, the presented theory is able to provide a coherent description
of the size effects of the Kondo resistivity in thin films, which is not
related to the size of the Kondo compensation cloud in any sense.

\section*{Acknowledgment}

The present authors are grateful for useful discussion with G. Bergmann,
L. Borda, N. Giordano, B. L. Gyorffy, H. v. L\"ohneysen, Ph. Nozi\`eres, 
M. Parpia, P. Phillips, J. S\'olyom, L. Szunyogh, I. K. Yanson and G. Zar\'and.
The work was supported by grants Hungarian OTKA T02228/1996 and T024005/1997.
One of us (O.\'U.) was supported by TEMPUS Mobility Grant and A.Z. is grateful 
for the support by the Humboldt Foundation.

\appendix

\section*{}
\label{app:1}

\begin{figure}
  \begin{center}
    \epsfig{file=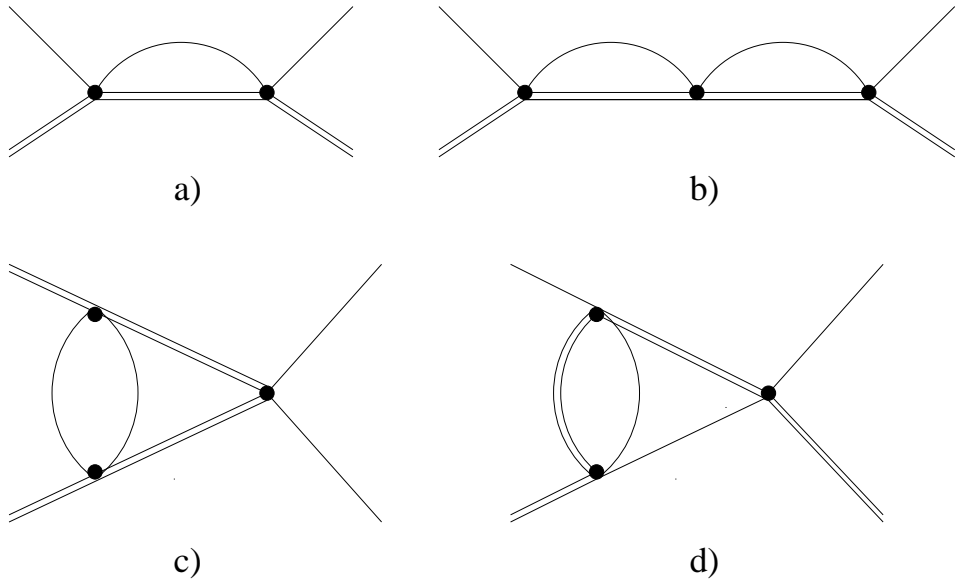}
  \end{center}
\caption{The second and third order vertex corrections.
The double line represents
the impurity spin with the anisotropy energy, the single one
the conduction electrons, and the solid circles stand for the exchange
interaction.}
\label{fig1}
\end{figure}

\begin{figure}
  \begin{center}
    \epsfig{file=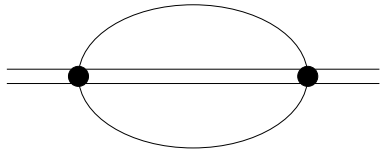}
  \end{center}
\caption{The second order self-energy correction for the impurity spin.
The double line represents
the impurity spin with the anisotropy energy, the single one
the conduction electrons, and the solid circles stand for the exchange
interaction.}
\label{fig3}
\end{figure}

\begin{figure}
  \begin{center}
    \epsfig{file=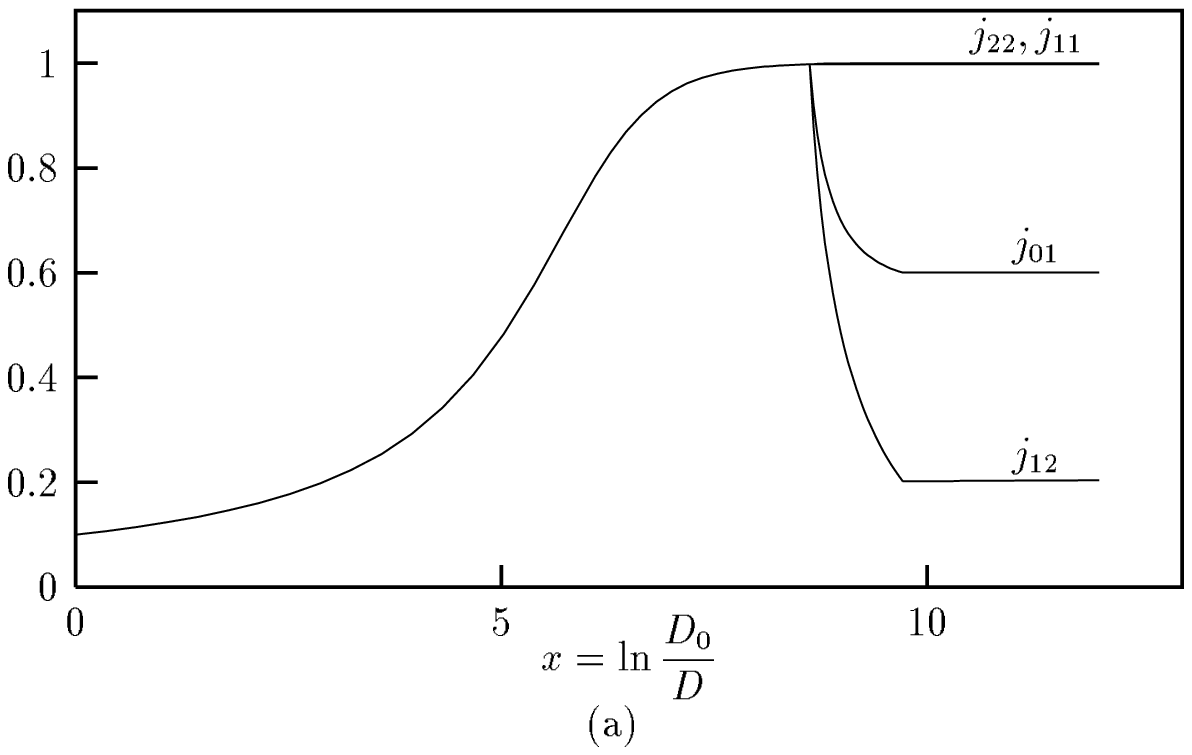}
    \epsfig{file=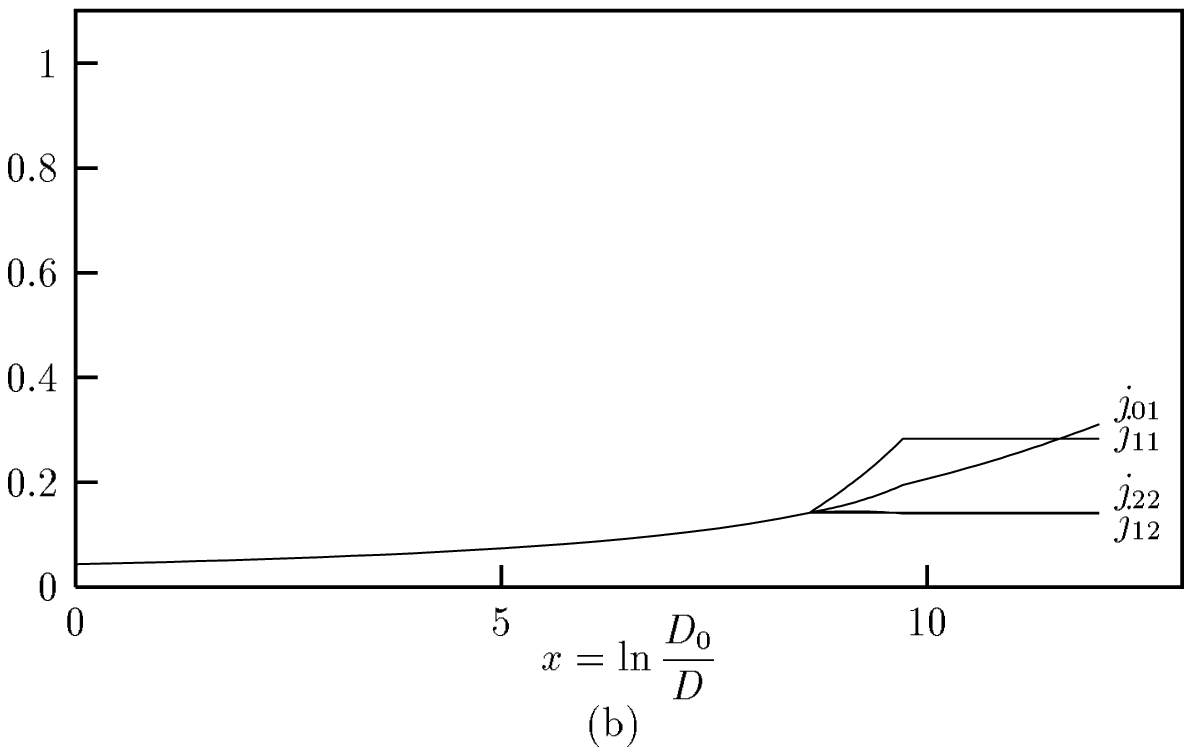}
  \end{center}
\caption{The running couplings for $S=2$ as a function of $x=\ln{D_0
\over D}$ at $K=6$ K, $T=0.6$ K with parameters $D_0=10^5 K$
and (a) $j_0=0.1$ (b) $j_0=0.0435$.}
\label{fig4}
\end{figure}

\begin{figure}
  \begin{center}
    \epsfig{file=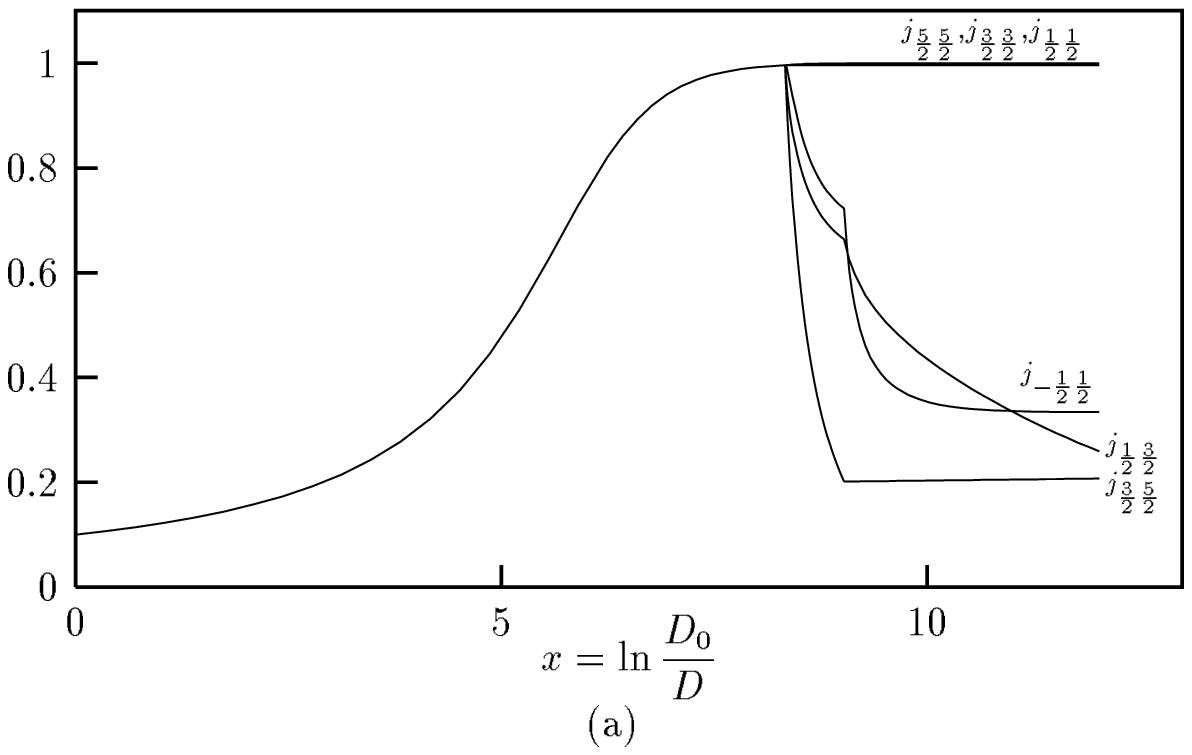}
    \epsfig{file=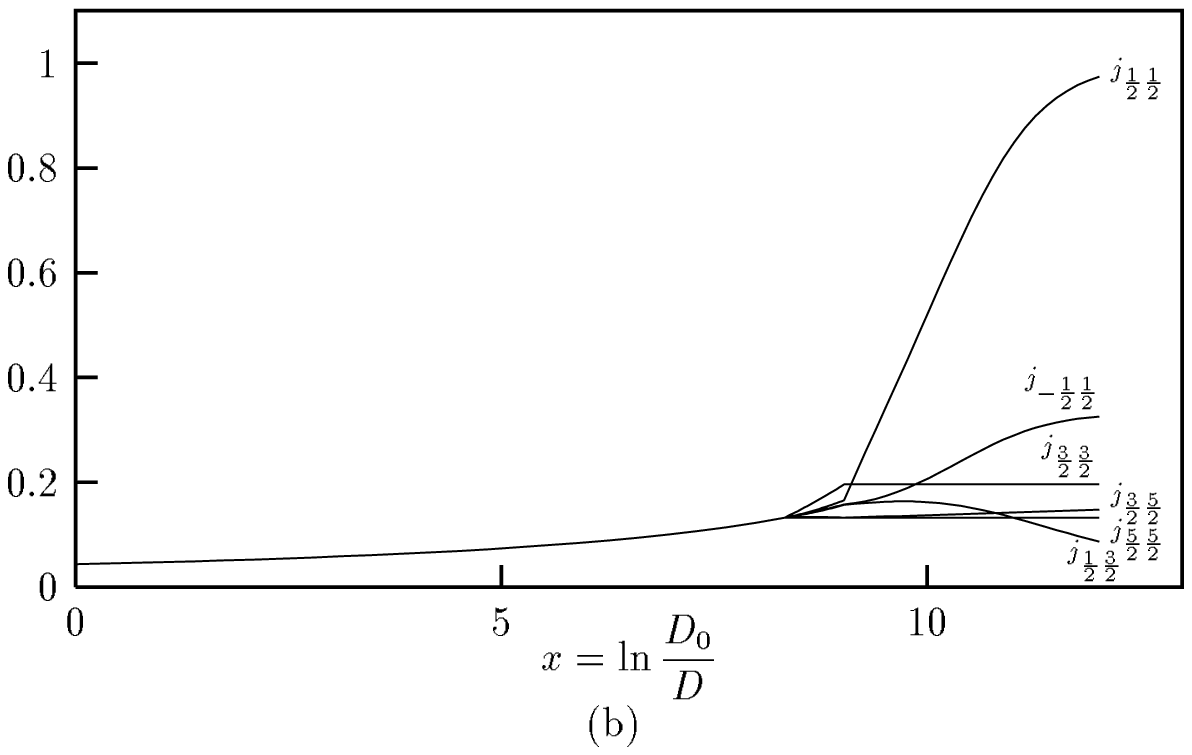}
  \end{center}
\caption{The running couplings for $S=5/2$ as a function of $x=\ln{D_0
\over D}$ at $K=6$ K, $T=0.6$ K with parameters $D_0=10^5$ K
and (a) $j_0=0.1$ (b) $j_0=0.0435$.}
\label{fig5}
\end{figure}

\begin{figure}
  \begin{center}
    \epsfig{file=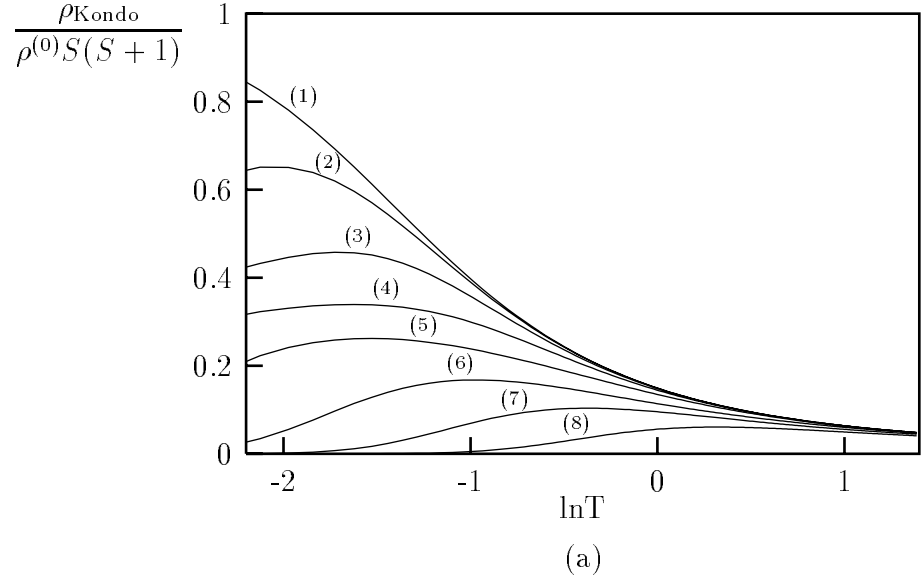}
    \epsfig{file=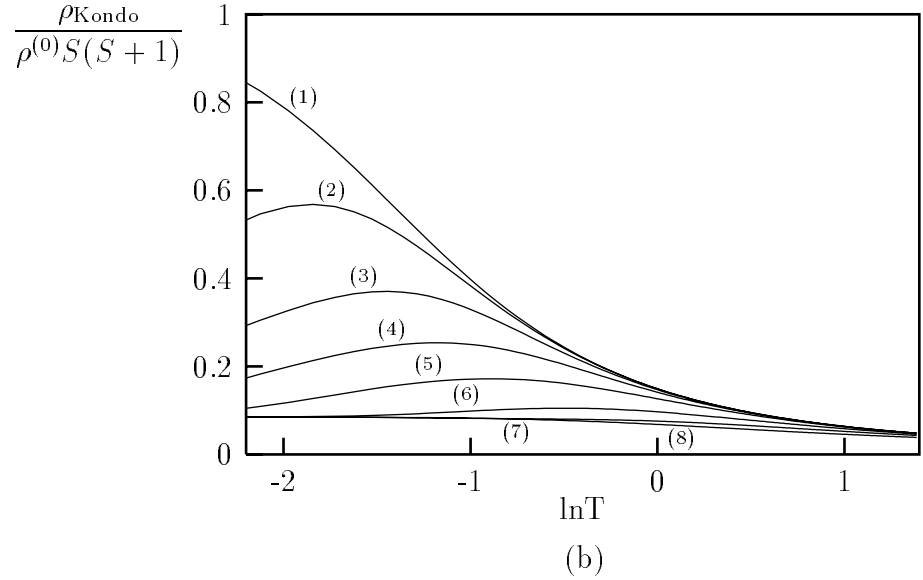}
  \end{center}
\caption{The resistivity for (a) $S=2$ and (b) $S=5/2$ for different values
of $K$. (1) $K=0$ (2) $K=0.02$ K (3) $K=0.05$ K (4) $K=0.1$ K (5) $K=0.2$ K
(6) $K=0.5$ K (7) $K=1$ K (8) $K=2$ K. The initial parameters were choosen
as $j_0=0.0435$ and $D_0=10^5$ K, $T_K=0.3$ K.}
\label{fig6}
\end{figure}

\begin{figure}
  \begin{center}
    \epsfig{file=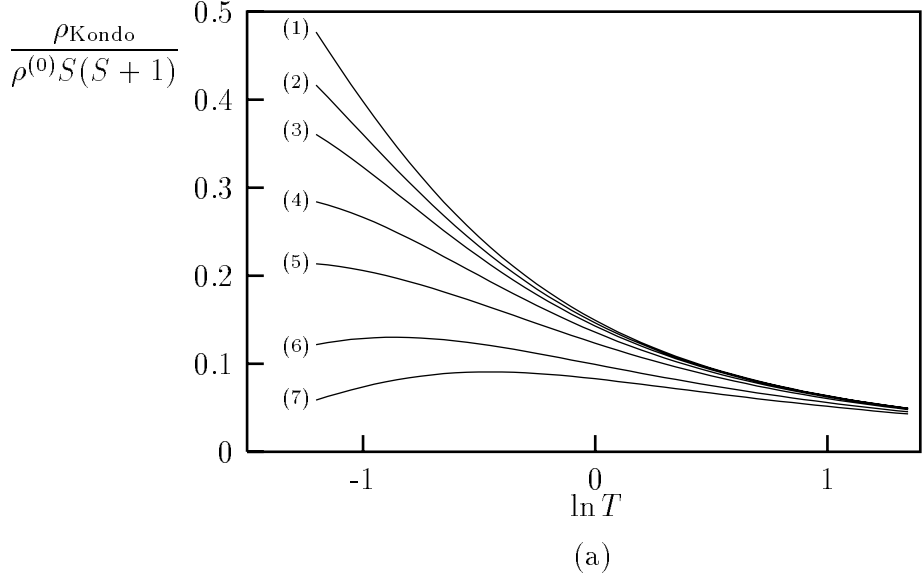}
    \epsfig{file=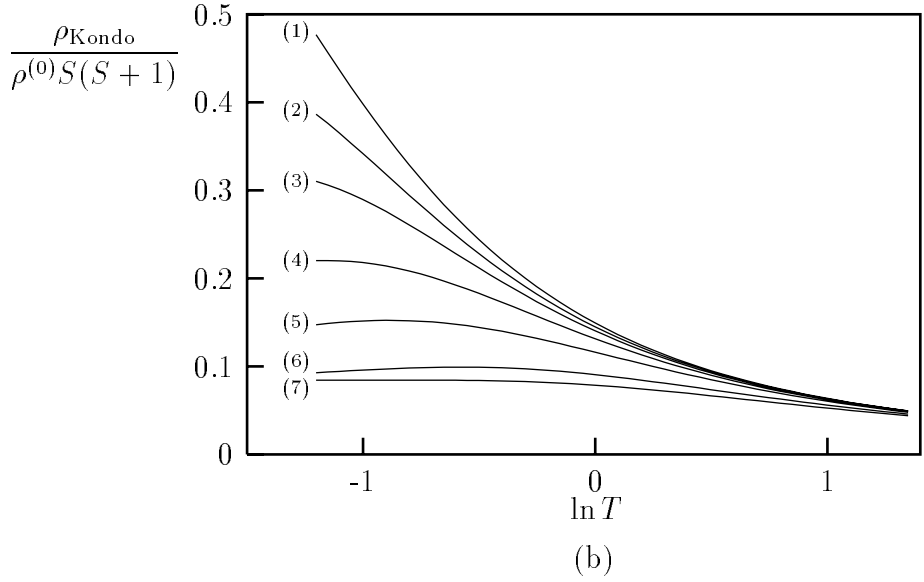}
  \end{center}
\caption{The resistivity for (a) $S=2$ and (b) $S=5/2$ for different values
of $t/\alpha$. (1) $t/\alpha=\infty (K=0)$ (2) $t/\alpha=200{1\over K}$
(3) $t/\alpha=100{1\over K}$ (4) $t/\alpha=50{1\over K}$
(5) $t/\alpha=25{1\over K}$ (6) $t/\alpha=10{1\over K}$
(7) $t/\alpha=6{1\over K}$.
The initial parameters were choosen
as $j_0=0.0435$ and $D_0=10^5$ K, $T_K=0.3$ K.}
\label{figatlagolt}
\end{figure}

\begin{figure}
  \begin{center}
    \epsfig{file=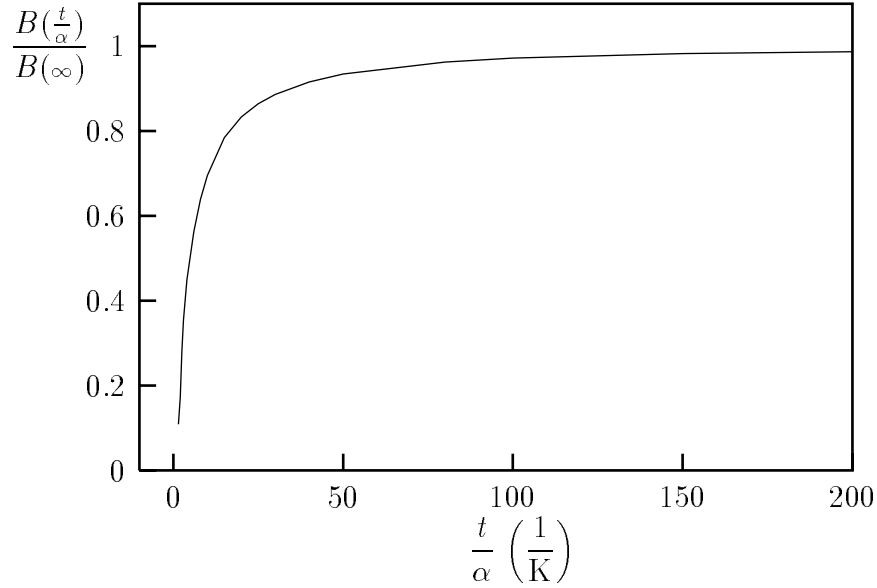}
  \end{center}
\caption{The calculated coeffitient $B$ as a function of $t/\alpha$.
The Kondo temperature was choosen as $T_K=0.3$ K.}
\label{fig10}
\end{figure}

\begin{figure}
  \begin{center}
    \epsfig{file=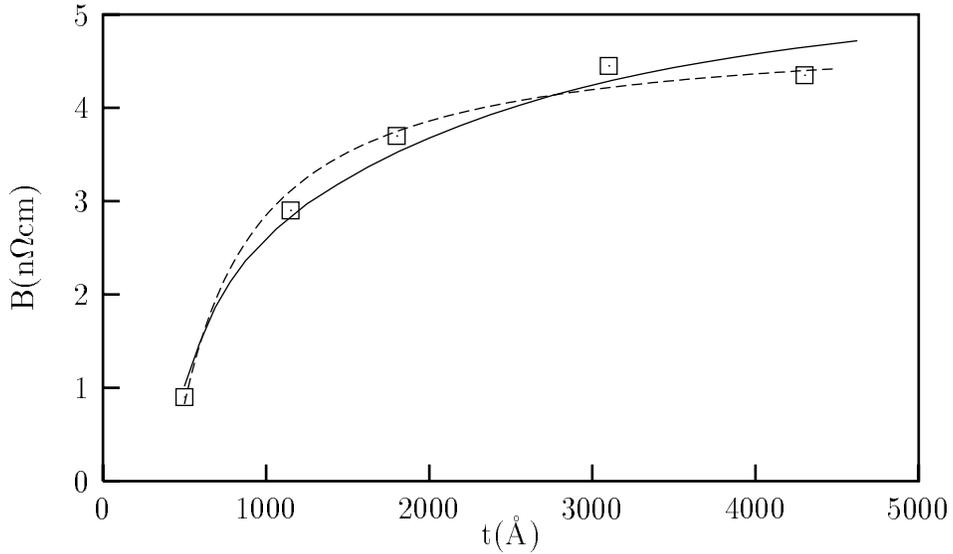}
  \end{center}
\caption{Fit on the experimental data (squares) by the calculated formula
$B(t)=\rho^{(0)} B_{\text{calc}}(t/\alpha)$ (the Kondo temperature
was choosen as $T_K=0.3$ K)
with fitting parameters
$\rho^{(0)}=20\text{n}\Omega\text{cm}$, $\alpha=247.7\text{\AA}$ K
(solid line) and by the phenomenological theory
$B(t)=B_{\infty} (1-2\lambda/t)$
with fitting parameters
$B_{\infty}=4.87 \text{n}\Omega\text{cm}$, $\lambda=207.5\text{\AA}$
(dashed line). The fit is not too sensitive to small changes ($<5\%$)
in $\alpha$.}
\label{kisfit}
\end{figure}

\begin{figure}
  \begin{center}
    \epsfig{file=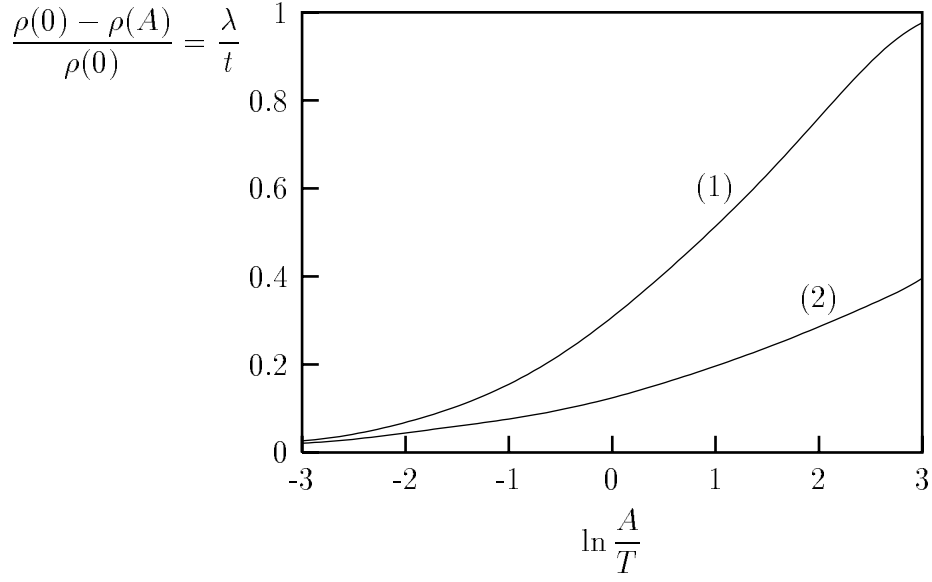}
  \end{center}
\caption{The effect of the mean free path on the Kondo resistivity in the
presence of anisotropy arising only one of the surface in
a thin film with thickness $L=600\text{\AA}$ for $S=2$ at $T=0.6$ K
(1) $K=A {\xi\over d}$ (2) $K=A {\xi\over d} e^{-d/\xi}$.
The Kondo temperature was choosen as $T_K=0.3$ K.}
\label{aperte}
\end{figure}

\begin{figure}
  \begin{center}
    \epsfig{file=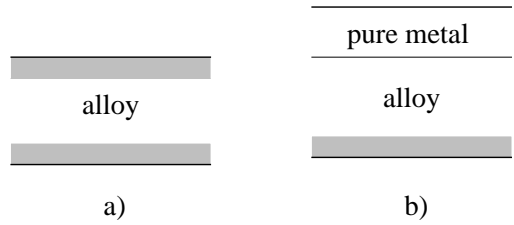}
  \end{center}
\caption{Bilayer structure}
\label{fig11}
\end{figure}

\begin{figure}
  \begin{center}
    \epsfig{file=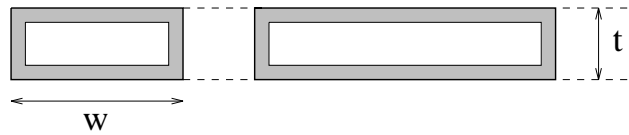}
  \end{center}
\caption{Stripes with same thickness $t$ and changing width $w$.}
\label{fig12}
\end{figure}

Here we calculate the second and third order vertex corrections,
and the second order self-energy correction for the impurity spin shown
in Fig.~\ref{fig1} and Fig.~\ref{fig3}, respectively.

Carrying out the Matsubara's summation, analytical continuation, changing 
the integrals $\int {d^3 k\over (2\pi)^3}$ to $\int\rho(\varepsilon)
d\varepsilon \int {d\Omega_k\over 4\pi}$ and using the assumption for $\rho(
\varepsilon)$ in Section~\ref{sec:1}, the contribution of the second
order diagrams are
\begin{eqnarray}
&&J_{MM^{''}}J_{M^{''}M'}(\sigma^i\sigma^j)_{\sigma\sigma'}
S^i_{MM^{''}}S^j_{M^{''}M'}\rho_0\int\limits_{-D}^D d\varepsilon
{1-n_F(\varepsilon)\over \varepsilon-\omega+K_{M^{''}}-K_M}\nonumber \\
+&&J_{MM^{''}}J_{M^{''}M'}(\sigma^j\sigma^i)_{\sigma\sigma'}
S^i_{MM^{''}}S^j_{M^{''}M'}\rho_0\int\limits_{-D}^D d\varepsilon
{n_F(\varepsilon)\over \varepsilon-\omega+K_{M'}-K_{M^{''}}},
\label{2vertexcorr}
\end{eqnarray}
for diagrams corresponding to Fig.\ \ref{fig1} (a).
The third order diagrams' contributions are
\begin{eqnarray}
-&&(\sigma^i\sigma^j\sigma^k)_{\sigma\sigma'} J_{MN} J_{NN'} J_{N'M'} 
S^i_{MN} S^j_{NN'} S^k_{N'M'}\rho_0^2\nonumber \\
&&\cdot\int\limits_{-D}^D 
{(1-n_F(\varepsilon) d\varepsilon)\over\omega-\varepsilon+K_M-K_N}
\int\limits_{-D}^D {(1-n_F(\varepsilon')) d\varepsilon'\over
\omega-\varepsilon'+K_M-K_{N'}}\nonumber \\ 
-&&(\sigma^k\sigma^j\sigma^i)_{\sigma\sigma'} J_{MN} J_{NN'} J_{N'M'}
S^i_{MN} S^j_{NN'} S^k_{N'M'}\rho_0^2\nonumber \\
&&\cdot\int\limits_{-D}^D 
{n_F(\varepsilon) d\varepsilon\over\varepsilon-\omega+K_{M'}-K_N}
\int\limits_{-D}^D {n_F(\varepsilon') d\varepsilon'\over
\varepsilon'-\omega+K_{M'}-K_{N'}}
\label{3vertexcorrb}\end{eqnarray}
for diagrams corresponding Fig.\ \ref{fig1} (b),
\begin{eqnarray}
-Tr(\sigma^i\sigma^j)\sigma^k_{\sigma\sigma'} J_{MN} J_{NN'} J_{N'M'}&&
S^i_{MN} S^j_{N'M'} S^k_{NN'}\rho_0^2\nonumber \\
&&\cdot\int\limits_{-D}^D 
\int\limits_{-D}^D {n_F(\varepsilon)(1-n_F(\varepsilon'))
d\varepsilon d\varepsilon'\over
(\varepsilon-\varepsilon'+K_M-K_N)(\varepsilon-\varepsilon'
+K_{M'}-K_{N'})}
\label{3vertexcorrc}\end{eqnarray}
for diagram Fig.\ \ref{fig1} (c), and
\begin{eqnarray}
&&(\sigma^i\sigma^j\sigma^k)_{\sigma\sigma'} J_{MN} J_{NN'} J_{N'M'} 
S^j_{MN} S^i_{NN'} S^k_{N'M'}\rho_0^2\nonumber \\
&&\cdot\int\limits_{-D}^D 
\int\limits_{-D}^D {n_F(\varepsilon)(1-n_F(\varepsilon'))
d\varepsilon d\varepsilon'\over
(\omega-\varepsilon'+K_M-K_{N'})(\varepsilon-\varepsilon'+K_M-K_N)}
\nonumber \\
+&&(\sigma^k\sigma^j\sigma^i)_{\sigma\sigma'} J_{MN} J_{NN'} J_{N'M'} 
S^j_{MN} S^i_{NN'} S^k_{N'M'}\rho_0^2\nonumber \\
&&\cdot\int\limits_{-D}^D 
\int\limits_{-D}^D {(1-n_F(\varepsilon))n_F(\varepsilon')
d\varepsilon d\varepsilon'\over
(\varepsilon'-\omega+K_{M'}-K_{N'})(\varepsilon'-\varepsilon+K_M-K_N)}
\nonumber \\
+&&(\sigma^i\sigma^j\sigma^k)_{\sigma\sigma'} J_{MN'} J_{N'N} J_{NM'} 
S^i_{MN} S^k_{NN'} S^j_{N'M'}\rho_0^2\nonumber \\
&&\cdot\int\limits_{-D}^D 
\int\limits_{-D}^D {n_F(\varepsilon)(1-n_F(\varepsilon'))
d\varepsilon d\varepsilon'\over
(\omega-\varepsilon'+K_M-K_N)(\varepsilon-\varepsilon'+K_{M'}-K_{N'})}
\nonumber \\
+&&(\sigma^k\sigma^j\sigma^i)_{\sigma\sigma'} J_{MN'} J_{N'N} J_{NM'} 
S^i_{MN} S^k_{NN'} S^j_{N'M'}\rho_0^2\nonumber \\
&&\cdot\int\limits_{-D}^D 
\int\limits_{-D}^D {(1-n_F(\varepsilon))n_F(\varepsilon')
d\varepsilon d\varepsilon'\over
(\varepsilon'-\omega+K_{M'}-K_N)(\varepsilon'-\varepsilon+K_{M'}-K_{N'})}
\label{3vertexcorrd}
\end{eqnarray}
for diagrams corresponding Fig.\ \ref{fig1} (d). 

The second order correction to the self-energy for the impurity spin
according to Fig.\ \ref{fig3} is
\begin{equation}
-Tr(\sigma^i\sigma^j) J_{MM'} J_{M'M} S^i_{MM'} S^j_{M'M}
\rho_0^2\int\limits_{-D}^D d\varepsilon\int\limits_{-D}^D d\varepsilon'
{(1-n_F(\varepsilon))n_F(\varepsilon')\over \varepsilon-\varepsilon'
-{\tilde\omega}+K_{M'}-K_M}.
\label{2selfencorr}\end{equation}
For the same indices summation must be carried out.

The spin factors in Eq.\ (\ref{2vertexcorr}), (\ref{3vertexcorrb}),
(\ref{3vertexcorrc}), (\ref{3vertexcorrd}) and
(\ref{2selfencorr}) were calculated by using the identities
\begin{mathletters}
\begin{equation}
(\sigma^i\sigma^j)_{\sigma\sigma'}
=\delta_{ij}\delta_{\sigma\sigma'}+i\varepsilon_{ijk}\sigma^k_{\sigma\sigma'},
\end{equation}
\begin{equation}
(\sigma^i\sigma^j\sigma^k)_{\sigma\sigma'}=\delta_{ij}\sigma^k_{\sigma\sigma'}
+\delta_{jk}\sigma^i_{\sigma\sigma'}-\delta_{ik}\sigma^j_{\sigma\sigma'}+
i\varepsilon_{ijk}\delta_{\sigma\sigma'},
\end{equation}
\end{mathletters}
introducing the $S^{\pm}$ operators in a usual way, and
exploiting that their matrix elements are
\begin{mathletters}
\label{Spm}
\begin{equation}
S_{MM'}^+=p(S,M')\delta_{M,M'+1}=\sqrt{S(S+1)-M'(M'+1)}\delta_{M,M'+1}
\end{equation}
\begin{equation}
S_{MM'}^-=q(S,M')\delta_{M,M'-1}=\sqrt{S(S+1)-M'(M'-1)}\delta_{M,M'-1}.
\end{equation}
\end{mathletters}

Turning to the integrals in Eq.\ (\ref{2vertexcorr}), (\ref{3vertexcorrb}),
(\ref{3vertexcorrc}), (\ref{3vertexcorrd}) and
(\ref{2selfencorr}), after changing the integration variable in integrals 
containing ($1-n_F(\varepsilon)$) or ($1-n_F(\varepsilon')$), from
$\varepsilon$ ($\varepsilon'$) to $-\varepsilon$ 
($-\varepsilon'$) they were evaluated in logarithmic approximation.
The integrals in Eq.\ (\ref{2vertexcorr}) give logarithmic contribution
\begin{eqnarray}
I^{(1)}_{MM''}(D)&=&\int\limits_{-D}^D d\varepsilon
{n_F(\varepsilon)\over \varepsilon+\omega+K_M-K_{M''}}\nonumber \\
\approx && \ln\left |{\omega+K_M-K_{M''}\over D}\right |+
I(\omega+K_M-K_{M''}), 
\label{type1}\end{eqnarray}
for $D>|\omega+K_M-K_{M''}|$ and
\begin{eqnarray}
I^{(2)}_{M'M''}(D)&=&\int\limits_{-D}^D d\varepsilon
{n_F(\varepsilon)\over \varepsilon-\omega+K_{M'}-K_{M''}}\nonumber \\
\approx && \ln\left |{\omega-K_{M'}+K_{M''}\over D}\right |+I(\omega-K_{M'}+
K_{M''}), 
\label{type2}\end{eqnarray}
for $D>|-\omega+K_{M'}-K_{M''}|$.

The integral in Eq.\ (\ref{2selfencorr}) gives logarithmic contribution
\begin{eqnarray}
I^{(3)}_{MM'}(D)&=&\int\limits_{-D}^D d\varepsilon\int\limits_{-D}^D 
d\varepsilon'{n_F(\varepsilon) n_F(\varepsilon')\over \varepsilon
+\varepsilon'+{\tilde\omega}-\lambda+K_M-K_{M'}}\nonumber \\
\approx && ({\tilde\omega}-\lambda+K_M-K_{M'})
\ln\left |{{\tilde\omega}-\lambda+K_M-K_{M'}\over D}\right |,
\label{type3}\end{eqnarray}
for $D>|{\tilde\omega}-\lambda+K_M-K_{M'}|$.

The integrals in Eq.\ (\ref{3vertexcorrb}) give logarithmic contribution
\begin{eqnarray}
I^{(4)}_{MNN'}(D)&=&I^{(1)}_{MN}(D)\,I^{(1)}_{MN'}(D)\nonumber \\
\approx && \biggl (\ln\left |{\omega+K_M-K_N\over D}\right |+I(\omega+K_M-K_N)
\biggr )\nonumber \\
&&\cdot \biggl (\ln\left |{\omega+K_M-K_{N'}\over D}\right |+
I(\omega+K_M-K_{N'})\biggr )
\label{type4}\end{eqnarray}
for $D>|\omega+K_M-K_N|$ and $D>|\omega+K_M-K_{N'}|$ and
\begin{eqnarray}
I^{(5)}_{M'NN'}(D)&=&I^{(2)}_{M'N}(D)\,I^{(2)}_{M'N'}(D)\nonumber \\
\approx && \biggl (\ln\left |{\omega-K_{M'}-K_N\over D}\right |+
I(-\omega+K_M-K_N)\biggr )\nonumber \\
&&\cdot\biggl (\ln\left |{\omega-K_{M'}-K_{N'}\over D}\right |+
I(-\omega+K_{M'}-K_{N'})
\biggr )
\label{type5}\end{eqnarray}
for $D>|-\omega+K_{M'}-K_N|$ and $D>|-\omega+K_M-K_{N'}|$.

The integral in Eq.\ (\ref{3vertexcorrc}) gives logarithmic contribution
for $M\neq N$, $M'\neq N'$ and $K_{M'}-K_{N'}-K_M+K_N\neq 0$
\begin{eqnarray}
I^{(6)}_{MNM'N'}(D)&=&\int\limits_{-D}^D d\varepsilon\int\limits_{-D}^D 
d\varepsilon'{n_F(\varepsilon) n_F(\varepsilon')\over 
(\varepsilon+\varepsilon'+K_M-K_N)(\varepsilon+\varepsilon'
+K_{M'}-K_{N'})}\nonumber \\
\approx && {K_M-K_N\over K_{M'}-K_{N'}-K_M+K_N}\ln\left |
{K_M-K_N \over D}\right |\nonumber \\
&-&{K_{M'}-K_{N'}\over K_{M'}-K_{N'}-K_M+K_N}
\ln\left |{K_{M'}-K_{N'}\over D}\right |
\label{type6a}\end{eqnarray}
for $D>|K_M-K_N|$, $D>|K_{M'}-K_{N'}|$.
For $K_{M'}-K_{N'}=K_M-K_N\neq 0$
\begin{equation}
I^{(6)}_{MNM'N'}(D)\approx -\ln\left |{K_M-K_N \over D}\right |,
\label{type6b}\end{equation}
for $D>|K_M-K_N|$.
And
\begin{equation}
I^{(6)}_{MMM'M'}(D)\approx -\ln\left |{T\over D}\right |.
\label{type6c}\end{equation} 
The integrals in Eq.\ (\ref{3vertexcorrd}) give logarithmic contribution
\begin{eqnarray}
I^{(7)}_{MN'MN}(D)&=&\int\limits_{-D}^D d\varepsilon\int\limits_{-D}^D 
d\varepsilon'{n_F(\varepsilon) n_F(\varepsilon')\over
(\varepsilon'+\omega+K_M-K_{N'})(\varepsilon+\varepsilon'+K_M-K_N)}
\nonumber \\
\approx && {1\over 2} \ln^2\left |{\omega+K_M-K_{N'}\over D}\right |+
I(\omega+K_M-K_{N'}) \ln\left |{\sqrt{(K_M-K_N)^2+T^2}\over D}\right |,
\label{type7}\end{eqnarray}
for $D>|\omega+K_M-K_{N'}|$ and $D>\sqrt{(K_M-K_N)^2+T^2}$,
where $T$ is the temperature,
\begin{eqnarray}
&&I^{(8)}_{M'N'MN}(D)=\int\limits_{-D}^D d\varepsilon\int\limits_{-D}^D 
d\varepsilon'{n_F(\varepsilon) n_F(\varepsilon')\over
(\varepsilon'-\omega+K_{M'}-K_{N'})(\varepsilon+\varepsilon'+K_M-K_N)}
\nonumber \\
\approx && {1\over 2} \ln^2\left |{-\omega+K_{M'}-K_{N'}\over D}\right |+
I(-\omega+K_{M'}-K_{N'}) \ln\left |{\sqrt{(K_M-K_N)^2+T^2}\over D}\right |,
\label{type8a}\end{eqnarray}
for $D>|-\omega+K_{M'}-K_{N'}|$ and $D>\sqrt{(K_M-K_N)^2+T^2}$,
\begin{eqnarray}
&&I^{(9)}_{MNM'N'}(D)=\int\limits_{-D}^D d\varepsilon\int\limits_{-D}^D
d\varepsilon'{n_F(\varepsilon) n_F(\varepsilon')\over
(\varepsilon'-\omega+K_{M}-K_{N})(\varepsilon+\varepsilon'+K_{M'}-K_{N'})}
\nonumber \\
=&&I^{(7)}_{MNM'N'}(D),
\end{eqnarray}
and
\begin{eqnarray}
&&I^{(10)}_{M'NM'N'}(D)=\int\limits_{-D}^D d\varepsilon\int\limits_{-D}^D
d\varepsilon'{n_F(\varepsilon) n_F(\varepsilon')\over
(\varepsilon'-\omega+K_{M'}-K_{N})(\varepsilon+\varepsilon'+K_{M'}-K_{N'})}
\nonumber \\
=&&I^{(8)}_{M'NM'N'}(D).
\end{eqnarray}

In the estimations above, the function $I(\alpha)=\int\limits_0^{\infty}
d\varepsilon n_F(\varepsilon) {2\varepsilon\over \varepsilon^2-\alpha^2}$ 
was introduced which is related to the finite $T$ divergences. From the 
scaling equations the $I(\alpha)$ function is canceled out.

For the sake of handling these contributions more comfortable
$\omega,{\tilde\omega}\sim T$ were set in the arguments of logarithms
in a way that substituting $\omega+a$ (${\tilde\omega}+a$) with 
$\sqrt{a^2+T^2}$.



\end{document}